\documentstyle[12pt,aaspp4]{article}

\begin{document}

\title{ISO Mid-Infrared Spectra of Reflection Nebulae$^{1}$}
\author{K. I. Uchida$^{2,3,4}$, K. Sellgren$^{2}$, M. W. Werner$^{5}$, 
        and M. L. Houdashelt$^{2,6}$}

\vskip 24pt

{\hspace*{24pt} Accepted for publication in the Astrophysical Journal, 5 October 1999}

\altaffiltext{1}
{Based on observations with ISO, an ESA project with instruments
funded by ESA Member States (especially the PI countries: France, 
Germany, the Netherlands and the United Kingdom) and with the 
participation of ISAS and NASA.}

\altaffiltext{2}
{Dept. of Astronomy, The Ohio State University, 140 West 
18$^{th}$ Ave., Columbus, OH 43210--1173}

\altaffiltext{3}
{kuchida@astrosun.tn.cornell.edu}

\altaffiltext{4}
{Current Address: Cornell University, 108 Space Sciences Bldg,
Ithaca, NY 14853 }

\altaffiltext{5}
{Jet Propulsion Laboratory, MS 264-767, 4800 Oak Grove Dr., Pasadena, 
CA 91109}

\altaffiltext{6}
{Current Address: Dept. of Physics \& Astronomy, Johns Hopkins University, 
3400 N. Charles St., Baltimore, MD 21218}
 
\begin{abstract}

We present 5 -- 15 $\mu$m imaging spectroscopy of the reflection
nebulae vdB\,17 (NGC 1333), vdB\,59 (NGC 2068), vdB\,101, vdB\,111,
vdB\,133 and vdB\,135, obtained with the infrared camera and circular
variable filter wheel on the {\it Infrared Space Observatory} ({\it
ISO}).  These nebulae are illuminated by stars with {$T_{\rm eff}$} =
3,600 K -- 19,000 K, implying ultraviolet (UV; $\lambda$ $<$ 400 nm)
to total stellar flux ratios of {$F_{(\lambda < 400 {\rm
nm})}$/$F_{total}$}\ = 0.01 -- 0.87.  We detect the infrared emission
features (IEFs) at 6.2, 7.7, 8.6, 11.3, and 12.7 \micron, broad
emission features at 6 -- 9 $\mu$m and 11 -- 13 $\mu$m, and 5 -- 15
$\mu$m continuum emission, from the interstellar medium in vdB\,17,
vdB\,59, and vdB\,133 ({$F_{(\lambda < 400 {\rm nm})}$/$F_{total}$}\ =
0.22 -- 0.87), and place upper limits on the emission from the
interstellar medium in vdB\,101, vdB\,111, and vdB\,135 ({$F_{(\lambda
< 400 {\rm nm})}$/$F_{total}$}\ = 0.01 -- 0.20).

Our goal is to test predictions of models attributing the IEFs to
polycyclic aromatic hydrocarbons (PAHs).  Interstellar models predict
PAHs change from singly ionized to neutral as the UV intensity, $G_0$,
decreases.  The ratio of PAH emission at 6 -- 10 $\mu$m to PAH
emission at 10 -- 14 $\mu$m is expected to be ten times higher in
ionized than neutral PAHs.

We observe no spectroscopic differences with varying $T_{\rm eff}$.
We analyze the spectra of vdB 17 and vdB 59 as a function of distance
from the star, to see how the spectra depend on $G_0$ within each
source.  The only quantitative difference we find is a {\it
broadening} of the 7.7 $\mu$m IEF at $G_0$ = 20 -- 60 within vdB 17.
We observe only a 40\% change in the 6 -- 10 $\mu$m to 10 -- 14 $\mu$m
flux ratio over $G_0$ = 20 to 6 $\times$ 10$^4$.

\end{abstract}


\section{Introduction}

The infrared emission features (IEFs) at 3.3, 6.2, 7.7, 8.6, 11.3 and
12.7 \micron, and their associated continuum, have been the subject of
intense study over the last decade.  The IEF strengths correlate with
the C/O ratio in planetary nebulae, suggesting a carbon-based carrier
(Roche \& Aitken 1986; Cohen et al.  1986, 1989).  Duley \& Williams
(1981) were the first to comment that some IEF wavelengths are
characteristic of the bending and stretching modes of various C--H and
C--C bonds in aromatic hydrocarbons.  There has since been a steady
accumulation of evidence, both laboratory and observational, in
support of this assertion.  The specific nature of the carrier(s),
however, is yet to be determined.  The IEFs have been attributed to
polycyclic aromatic hydrocarbon (PAH) molecules (L\'eger \& Puget
1984; Allamandola, Tielens, \& Barker 1985) and to more amorphous
materials containing aromatic hydrocarbons (Sakata et al.  1984, 1987;
Borghesi, Bussoletti, \& Colangeli 1987; Blanco, Bussoletti, \&
Colangeli 1988; Duley 1988; Papoular et al.  1989).  Allamandola,
Tielens, \& Barker (1989) and Bregman et al.  (1989) attribute the
broad spectral structures at 6 -- 9 \micron\ and 11 -- 13 \micron,
which accompany the IEFs, to PAH clusters and amorphous carbon grains.

The origin of the 1 -- 25 $\mu$m continuum emission, like that of the
associated IEF emission, is still the subject of debate.  The
continuum is observed in regions of very low radiation density, such
as reflection nebulae and infrared cirrus, where the grain equilibrium
temperature is too low to explain the emission as being thermal in
nature (Sellgren 1984; Draine \& Anderson 1985; Guillois et al.
1996).  It has been attributed to single ultraviolet (UV) photon
excitation of very small grains (Sellgren, Werner, \& Dinerstein 1983;
Sellgren 1984), to PAH molecular fluorescence (L\'eger \& Puget 1984;
Allamandola et al.  1985, 1989; Puget \& L\'eger 1989), and to
hydrogenated amorphous carbon grain luminescence (Duley \& Williams
1988; Duley 1988).

The IEF carriers are a ubiquitous component of the interstellar medium
(ISM).  Past studies have found IEFs in {\it emission} near Galactic
\ion{H}{2} regions, reflection nebulae, planetary nebulae,
proto-planetary nebulae, and the diffuse ISM of our own and other
galaxies (see reviews by Aitken 1981; Allamandola et al.  1989; Puget
\& L\'eger 1989; Geballe 1997; and Tokunaga 1997).  The 6.2 $\mu$m IEF
has recently been detected in {\it absorption} in the diffuse ISM
(Schutte et al.  1996, 1998).  The 3.3 $\mu$m IEF may have also been
detected in {\it absorption} in molecular clouds (Sellgren, Smith, \&
Brooke 1994; Sellgren et al.  1995; Brooke, Sellgren, \& Smith 1996;
Brooke, Sellgren \& Geballe 1999).

IEF spectra have been recently classified into several groups (Geballe
1997; Tokunaga 1997).  Class A spectra at 5 -- 15 $\mu$m display
narrow IEFs at 6.2, 7.7, 8.6, 11.3 and 12.7 \micron, as well as
associated continuum emission.  The 7.7 \micron\ IEF is invariably the
strongest of the Class A narrow IEFs (Geballe 1997).  By contrast,
Class B spectra lack the 8.6 \micron\ IEF and have notably broader
features at 8 and 11.5 \micron\ (Kwok, Volk, \& Hrivnak 1989).  Class
B spectra sometimes also show a strong broad emission feature at 8.8
\micron\ (Buss et al.  1993).  The overwhelming majority of IEF
spectra currently measured fall under the category of Class A.
Sources with Class A IEFs are typically UV-rich sources, such as the
diffuse ISM, reflection nebulae, \ion{H}{2} regions, young stars,
planetary nebulae and galaxies hosting star formation.  By contrast,
the half-dozen or so Class B spectra found to date arise from F and
G-type post-AGB stars forming planetary nebulae (i.e.  proto-planetary
nebulae; Geballe et al.  1992).

Well-developed theoretical work exists for the absorption and emission
properties of PAHs, inspired by the hypothesis that PAHs are
responsible for the IEFs.  Mixtures of PAHs and other dust components
have been used to model the UV and visible extinction curve and the
infrared emission observed in the ISM (Puget, L\'{e}ger, \& Boulanger
1985; Chlewicki \& Laureijs 1988; D\'esert, Boulanger \& Puget 1990;
Joblin, L\'eger, \& Martin 1992; Siebenmorgen \& Kr\"ugel 1992;
Schutte, Tielens, \& Allamandola 1993; Dwek et al.  1997; Li \&
Greenberg 1997; Silva et al.  1998).  All of these models include a
size distribution of PAHs.  Most of these size distributions include
large PAHs, expected in the ISM but not yet measured in the
laboratory, by extrapolation from laboratory data on smaller PAHs.
Interstellar PAHs are predicted to be mainly positively charged and
neutral in strong UV fields, and mainly neutral and negatively charged
when the UV field is weak (Bakes \& Tielens 1994, 1998; Salama et al.
1996; Dartois and d'Hendecourt 1997).  Some interstellar models
(particularly D\'esert et al.  1990) include both neutral and ionized
PAHs.

These models for the absorption and emission from a size distribution
of neutral and ionized PAHs predict that the fraction of the total
interstellar extinction curve due to PAH absorption is much larger at
UV wavelengths than at visible wavelengths.  This is vividly
illustrated in Figures 2 and 3 of D\'esert et al.  (1990), Figures 1,
2, and 3 of Joblin et al.  (1992), Figures 1{\it b}, 2{\it b} and
4{\it b} of Siebenmorgen \& Kr\"ugel (1992), Figure 4 of Dwek et al.
(1997), and Figure 3 of Silva et al.  (1998).  If PAHs contribute more
to the total interstellar extinction at UV than visible wavelengths,
then PAHs should absorb and re-emit a larger fraction of the total
absorption and re-emission by dust in UV-rich sources than in UV-poor
sources.

Reflection nebulae are valuable tools for the study of the Class A
interstellar IEFs.  The localized heating of an interstellar cloud by
a nearby, optically visible, star allows us to study the IEFs over
varying excitation conditions by observing reflection nebulae
illuminated by stars of different effective temperature, $T_{\rm
eff}$.  This localized heating also results in brighter infrared
emission than would be observed in a similar cloud heated only by the
diffuse interstellar radiation field.

Sellgren, Luan, \& Werner (1990) analyzed {\it IRAS} observations of
reflection nebulae illuminated by stars with $T_{\rm eff}$ = 3,000 K
-- 33,000 K.  They observed that the ratio of nebular emission within
the {\it IRAS} 12 $\mu$m broadband filter, $\Delta \nu I_\nu$(12
$\mu$m), to the total bolometric far-infrared nebular emission,
$I_{\rm bol}$(FIR), was independent of $T_{\rm eff}$, for $T_{\rm
eff}$ = 5,000 K -- 33,000 K.  The total infrared emission of a
reflection nebula is due to starlight absorbed and re-radiated by the
mixture of molecules and grains responsible for the total interstellar
extinction curve.  Their observation that $\Delta \nu I_\nu$(12
$\mu$m)/$I_{\rm bol}$(FIR) is constant over a wide range of $T_{\rm
eff}$ implies that the material emitting in the {\it IRAS} 12 $\mu$m
band has an absorption curve with a wavelength dependence similar to
that of the total interstellar absorption curve over both UV and
visible wavelengths.

The Sellgren et al.  (1990) observations, finding $\Delta \nu
I_\nu$(12 $\mu$m)/$I_{\rm bol}$(FIR) to be independent of $T_{\rm
eff}$, appear to be in conflict with theoretical predictions that the
{\it IRAS} 12 $\mu$m band is dominated by PAH emission and that PAHs
contribute more to the total interstellar extinction curve at UV than
visible wavelengths (D\'esert et al.  1990; Joblin et al.  1992;
Siebenmorgen \& Kr\"ugel 1992; Schutte et al.  1993; Dwek et al.
1997; Li \& Greenberg 1997; Silva et al.  1998).  This is particularly
puzzling as both D\'esert et al.  (1990) and Siebenmorgen \& Kr\"ugel
(1992) explicitly fit a mixture of PAHs and other dust components to
the observed extinction curves and infrared emission of two reflection
nebulae observed by Sellgren et al.  (1990):  NGC 2023 (vdB 52;
$T_{\rm eff}$ = 22,000 K) and NGC 7023 (vdB 139; $T_{\rm eff}$ =
17,000 K).

The Sellgren et al.  (1990) observations could be reconciled with
theoretical predictions if other ISM components besides PAHs
contribute to the {\it IRAS} 12 $\mu$m broad-band emission in
reflection nebulae illuminated by cool stars.  Infrared spectroscopy
can test whether the fractional contribution of PAHs to the {\it IRAS}
12 $\mu$m broad-band emission of reflection nebulae depends on $T_{\rm
eff}$.  Sellgren, Werner, \& Allamandola (1996) have searched for
spectral differences in the near-infrared emission of reflection
nebulae with $T_{\rm eff}$ = 3,600 K -- 33,000 K.  They find that the
equivalent width of the 3.3 $\mu$m IEF in reflection nebulae is
independent of $T_{\rm eff}$, over $T_{\rm eff}$ = 11,000 K -- 22,000
K, but they did not have the sensitivity to detect reflection nebulae
illuminated by cooler stars.

This paper presents a spectroscopic imaging study to search for and
detail interstellar IEF emission toward a sample of reflection nebulae
illuminated by stars with $T_{\rm eff}$ = 3,600 K -- 19,000 K (Table
\ref{tab:stelpar}).  We obtained our observations with the {\it
Infrared Space Observatory} ({\it ISO}) camera, ISOCAM (Kessler et al.
1996; C.  Cesarsky et al.  1996), using the circular variable filter
(CVF).  We have reported our initial result, the detection of IEFs
toward the UV-poor nebula vdB\,133 (illuminated by two stars with
$T_{\rm eff}$ = 6,800 K and 12,000 K), in Uchida, Sellgren, \& Werner
(1998; hereafter Paper I).  The goal of our study is to determine
whether there are quantitative differences between the excitation of
the different IEFs and the continuum emission among nebulae with
widely varying $T_{\rm eff}$.  We selected reflection nebulae with
previous {\it IRAS} detections of extended 12 \micron\ emission
(Sellgren et al.  1990).  The unprecedented sensitivity of the
cryogenically cooled {\it ISO} satellite to low surface brightness
mid-infrared emission has allowed us to search for IEFs toward the
reflection nebulae illuminated by stars later than B-type, where the
IEFs are too faint for detection with ground-based or airborne
telescopes at ambient temperature.

\section{Observations}

\subsection{Position-Switched Data}

We obtained ISOCAM multi-wavelength images of vdB\,101, vdB\,111,
vdB\,133 and vdB\,135 between 1996 March\,19 and 1998 February 16.  We
used a set of 9 -- 10 CVF wavelengths and 1 -- 2 narrowband filters
toward each nebula, in four sequences.  Each sequence contains
observations at three to four wavelengths, as listed in Table
\ref{tab:obswave}.  We observed all four sequences for each nebula,
except for vdB\,135, where no Sequence 3 observations were made.
While the filter observations within a given sequence were performed
consecutively in time, the scheduling of the sequences themselves were
not necessarily constrained so.  Some sequences thus share common
wavelength observations to provide a check for consistency with time.
All observations were made with a 6\arcsec\ pixel$^{-1}$ scale giving
a total field of view of 3\farcm2 with the 32$\times$32 pixel array.
The CVF spectral resolution was $R$ = $\lambda/\Delta \lambda$ = 40
and the narrowband filters at 4.5 \micron\ (LW1) and 15 \micron\ (LW9)
had bandpasses of $\Delta\lambda$ = 1 \micron\ and 2 \micron,
respectively (ISOCAM Team 1994).

The observations were position-switched (in a 2-cycle
sky$\,\rightarrow\,$source$\,\rightarrow\,$sky$\,\rightarrow\,$source
sequence), with equal times on the source and sky to allow for the
removal of the background and dark current contributions.  The source
observations were centered on the illuminating star(s) of each nebula
and the sky positions were toward nearby regions deemed free, based on
inspection of the {\it IRAS} 12 \micron\ skyview images, of any
significant emission above the Zodiacal component and the local cirrus
background.  Table \ref{tab:obspos} lists the positions we observed.
The basic integration time per exposure varied between 0.28 s and 10
s, depending on the wavelength and brightness of the central star.
The total number of exposures was set to attain a total on-source
integration time per filter of $\sim$200\,s for all observations.

\subsection{Full CVF Wavelength Scans}

Spectral images of vdB\,17 (NGC 1333), vdB\,59 (NGC 2068), vdB\,101,
vdB\,133 and vdB\,135 were also taken during the supplemental
observing period of {\it ISO}.  These involved full CVF wavelength
scans at 5.14 -- 9.44 \micron\ (CVF1) and 9.33 -- 15.1 \micron\
(CVF2), taken in both the increasing and decreasing wavelength
directions, toward each of the source and sky positions.  The spectral
resolution was $R$ = 40.  The observations were spaced in wavelength
to provide a factor of 1 -- 2 oversampling in the vdB\,17 and vdB\,59
spectra, and 2 -- 4 oversampling in those of the other sources.  The
basic exposure time used was 2.1\,s, with 8 exposures per CVF step
(totalling 17.6\,s) toward the bright sources vdB\,17 and vdB\,59, and
11 exposures per CVF step (totalling 23.1\,s) toward vdB\,101,
vdB\,133 and vdB\,135.  The pixel scale was 6\arcsec\ pixel$^{-1}$ for
the vdB\,17, vdB\,59 and vdB\,133 measurements, like that of the
earlier position-switched observations, and 12\arcsec\ pixel$^{-1}$
for vdB\,101 and vdB\,135.  In this second set of observations, the
``source'' fields were offset from the central star so as not to
contain the star, thus allowing us to use longer integration times
without saturating the detector.  The vdB\,133 field was centered
toward the peak IEF emission in the region, as inferred from the
earlier position-switched {\it ISO} observations, and those of vdB\,17
and vdB\,59 were centered in the directions containing the strongest
extended infrared emission in our unpublished near-infrared
ground-based images of the regions.  The vdB\,101 and vdB\,135 fields
were centered on the same positions observed by Sellgren et al.
(1990) in their {\it IRAS} observations.  The sky positions were
chosen by examination of the 12 \micron\ {\it IRAS} skyflux images.
Table \ref{tab:obspos} lists the source and sky positions.

\section{Data Reduction}

The images were reduced, from the basic (CISP) data product, in the
IDL-based ISOCAM data reduction environment known as ``ICE'',
developed at the Institut d'Astrophysique Spatiale (IAS).  By
utilizing the basic ICE modules we were able to more closely monitor
and interact with the various stages of processing (i.e., the
deglitching of cosmic ray hits from the data, detector transient
response rectification and flat-fielding) than with the standard (more
automated) CAM IA processing package.  Indeed, special efforts were
made in the restoration of the source fluxes; the description of the
method follows.

\subsection{The ISO Camera Transient Response}

The ISOCAM detector array experiences a response lag and a
flux-dependent transient response (or residual memory effect).  In
order to recover the actual source fluxes, $I(t)$, models of the
detector behavior have been developed by several of the institutes
involved with ISOCAM.  We have adopted the transient model developed
at the IAS (Abergel et al.  1996),

\begin{equation}
       M(t) = r \ I(t) \ + \ (1-r) \ 
       \int ^ t _ {- \infty}  { I(t') \exp \left(\frac{t'-t}{\tau(t')}
       \right)\frac{dt'}{\tau(t')}} 
\end{equation}

\parindent0pt
which has proven from past experience to work best in rectifying
low-level flux like that expected of the diffuse nebular emission.
The measured flux, $M$, at time, $t$, is the sum of the instantaneous
response of the detector, $r \ I(t)$, and the integrated contribution
of the exponentially decaying flux history of the detector (Abergel et
al 1996).  The detector is characterized by two parameters, $r$ and
$\alpha$, the exact behavior of which is not well-known but which
likely varies somewhat from pixel to pixel and perhaps with time, over
the long term (weeks to months).  The time constant for the
exponential decay, $\tau\ = {\alpha}/{M(t)}$, is a function both of
the measured flux and of the detector parameter, $\alpha$.  In the
best of circumstances, the flux history of the detector prior to the
start of one's own observations is known and can then be used to
rectify the first part of the flux time series.  Since, however, the
data prior to one's own are generally not available, a number of
``padding'' frames, all with an estimated constant flux value, are
instead prepended.

\begin{figure}
\figurenum{1}
\plotone{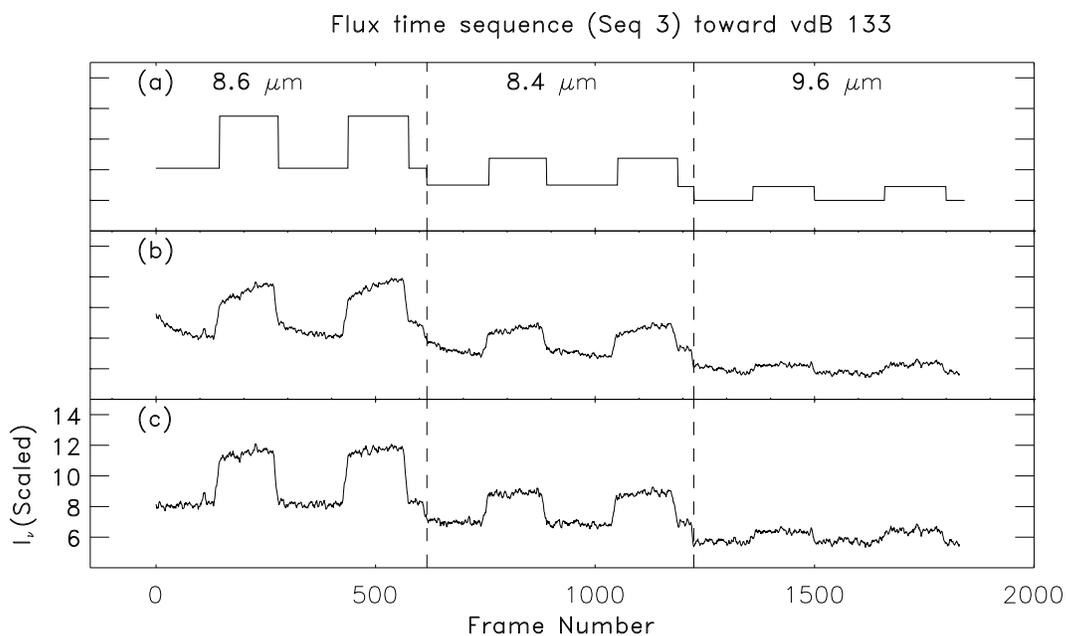}
\figcaption[]{Rectification of ISOCAM data for Sequence 3 for vdB 133,
during a 2 cycle sky$\,\rightarrow\,$source observational sequence
done consecutively with three different circular variable filter (CVF)
settings at wavelengths of 8.6, 8.4, and 9.6 $\mu$m.  The horizontal
axis is image frame number; each image was obtained with an
integration time of 0.28 s.  {\it a}:  the ideal response of the
detector.  {\it b}:  an actual observation sequence toward vdB\,133
showing the deviation from the ideal due to the detector flux
transient response.  {\it c}:  the data of panel {\it b} after
application of the flux ``rectification'' procedure.  \label{ts} }
\end{figure}

\parindent14pt
If the detector response were perfect (Figure \ref{ts}{\it a}), the
flux history of a 2-cycle chop
(sky$\,\rightarrow\,$source$\,\rightarrow\,$sky$\,\rightarrow\,$source)
sequence, for a single pixel or an average of pixels in the array,
should appear as a series of square functions with the data points
falling on two plateaus at a common minimum and two plateaus at a
common maximum; the entire pattern then repeats itself (but likely
with some general flux offset with respect to the first) for each of
the subsequent position-switched cycles taken with a different filter.
Figure \ref{ts}{\it b} displays an observed flux time sequence in raw
form.  The effects of the detector's finite response time and flux
memory are seen to cause significant deviations from the ideal square
pattern.  The individual source and sky segments of the flux sequence,
which should otherwise have constant levels, are either temporally
increasing or decreasing (depending on the history of the detected
signal) because of the non-ideal detector behavior.

The position-switched data lend themselves well to the determination
of the detector parameters.  By using the ideal square-wave pattern as
the target in a chi-square minimization procedure, with $r$, $\alpha$,
and the ``padding'' constant as free parameters, the best values of
these parameters that rectify the observed time sequences can be
found.  Indeed, a set of parameters was determined for each of the
four observation sequences toward vdB\,133 from averages of those
pixels containing the nebular emission.  The four parameter sets
ranged between $r$ = 0.72 -- 0.73 and $\alpha$\ = 954 -- 1603 ADU
gain$^{-1}$, with padding values of 9.2 -- 13.8 ADU gain$^{-1}$, where
ADU gain$^{-1}$ are analog-to-digital units divided by the detector
gain.  The specific parameter sets were used to rectify the four
corresponding vdB\,133 observation sequences, and their average ($r$ =
0.72 and $\alpha$\ = 1200 ADU gain$^{-1}$) was used to rectify the
data sequences of the other sources.  The parameter values determined
by us provided some improvement over the results based on the values
adopted by Abergel et al.  (1996), $r$ = 0.63 and $\alpha$\ = 1200 ADU
gain$^{-1}$, for observations toward $\rho$ Oph.  Figure \ref{ts}{\it
c} displays a rectified flux time sequence, after the observed flux
time sequence (Fig.  \ref{ts}{\it b}) has been rectified using the
appropriate parameter values.

In contrast, it is not possible to deduce the detector parameters from
the full CVF scans with the method described above, since these scans
involve sequential wavelength observations toward a given position and
thus do not possess a predictable pattern, much less the regular
``square wave'' pattern of the position-switched sequences.  All full
CVF scan sequences of our supplemental observations were thus
rectified using the average of the parameters determined from the
original vdB\,133 position-switched sequences:  $r$ = 0.72 and
$\alpha$\ = 1200 ADU gain$^{-1}$.

The source and sky images were rectified separately before being used
to produce the final source\,$-$\,sky maps.  In each case, the dark
current image (Boulanger 1996) was subtracted from the maps to
establish the absolute count levels required for the rectification
procedure.  It was discovered, however, that the odd-even noise
pattern (ISOCAM Team 1994) remained in the images after a simple
subtraction of the dark and that multiplication of the dark field
(with factors ranging over 1.03 -- 1.13) was needed to eliminate this
problem.  A single dark multiplication factor was used for all
observations of a given source.  An uncertainty of $\sim$0.03 ADU
gain$^{-1}$ in determining the dark factor value translates to an
uncertainty of $\sim$3 MJy sr$^{-1}$ in the individual source and sky
map levels.  However, since the source image and the sky image of a
given nebula each use the same dark field (with the same dark
multiplicative factor), any errors in absolute flux levels of the
component fields do not propagate into the source\,$-$\,sky maps.  The
level offset in the source and sky images may lead to differential
errors by the rectification process that appear in the
source\,$-$\,sky maps, however, on the order of 5\% of the absolute
sky level (\S4).

\section{Results}

\begin{figure}
\figurenum{2}
\plotone{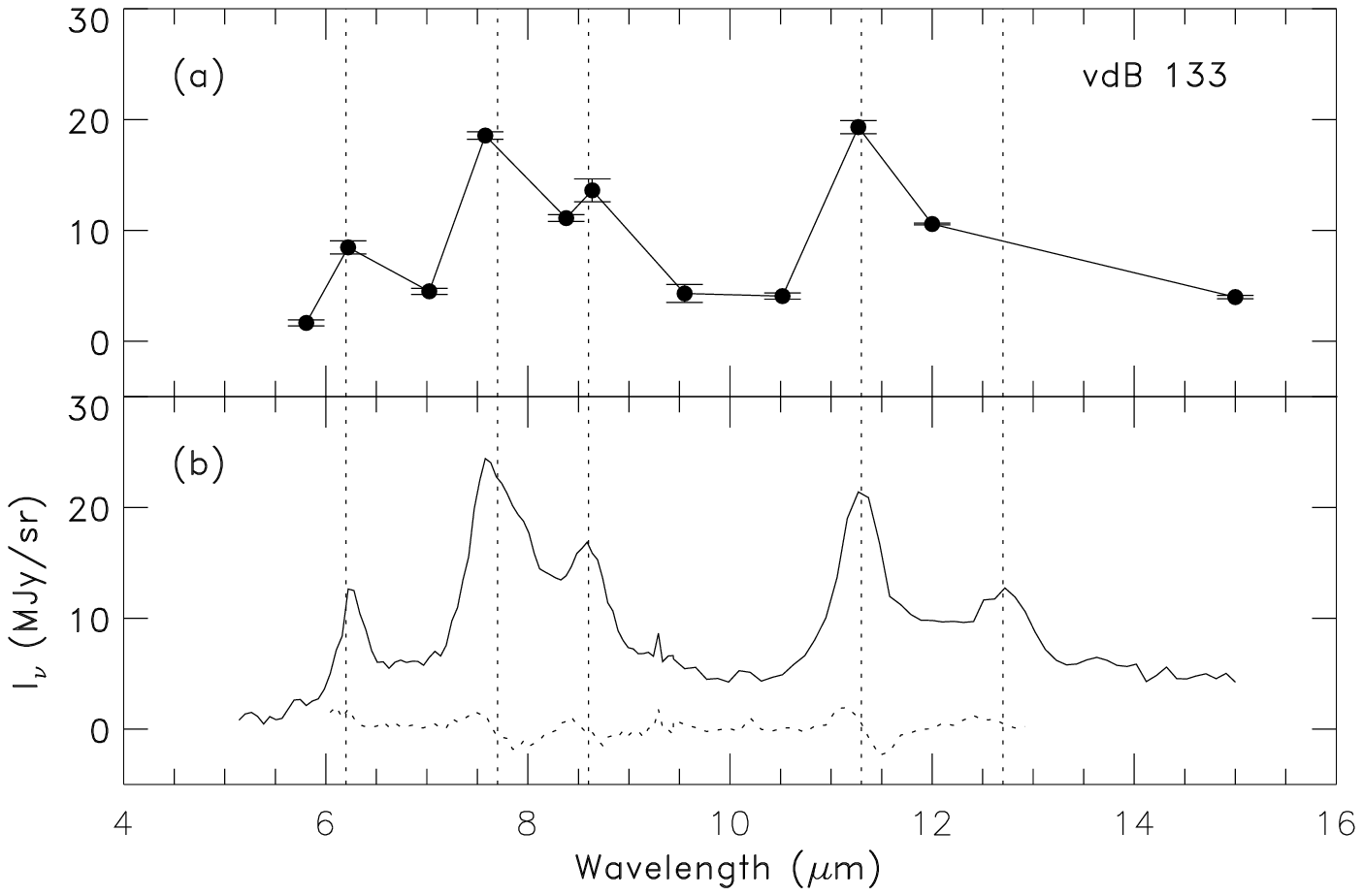}
\figcaption[]{Flux-rectified and sky-subtracted 5.1 -- 15 \micron\
spectra of vdB\,133, in $I_{\nu}$ (MJy sr$^{-1}$) vs.  $\lambda
(\mu$m).  Wavelengths of the 6.2, 7.7, 8.6, 11.3, and 12.7 \micron\
infrared emission features (IEFs) are marked ({\it dotted vertical
lines}).  ({\it a}) Spectrum from position-switched observations at
chosen CVF wavelengths ({\it filled circles}).  Error bars show the
difference between the cycle 1 and cycle 2 source$-$sky rectified data
at each wavelength.  ({\it b}) Spectrum from complete CVF scan ({\it
solid line}).  The difference between the rectified increasing and
decreasing wavelength scans of the source-sky spectra ({\it dotted
line}) is plotted as an estimate of the uncertainty.  \label{iso1} }
\end{figure}

\begin{figure}
\figurenum{3}
\plotone{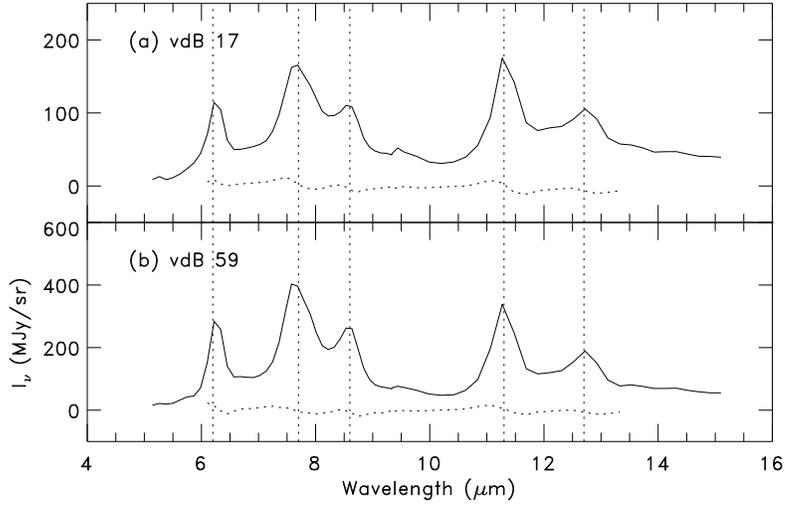}
\figcaption[]{Flux-rectified, sky-subtracted 5.1 -- 15 \micron\
spectra ({\it solid line}) from complete CVF scans of ({\it a})
vdB\,17 (NGC\,1333), and ({\it b}) vdB\,59 (NGC\,2068), in $I_{\nu}$
(MJy sr$^{-1}$) vs.  $\lambda (\mu$m).  The wavelengths of the 6.2,
7.7, 8.6, 11.3, and 12.7 \micron\ IEFs are marked ({\it dotted
vertical lines}).  The difference between the rectified increasing and
decreasing wavelength scans of the source-sky spectra ({\it dotted
line}) is plotted as an estimate of the uncertainty.  \label{iso2} }
\end{figure}

\begin{figure}
\figurenum{4}
\plotone{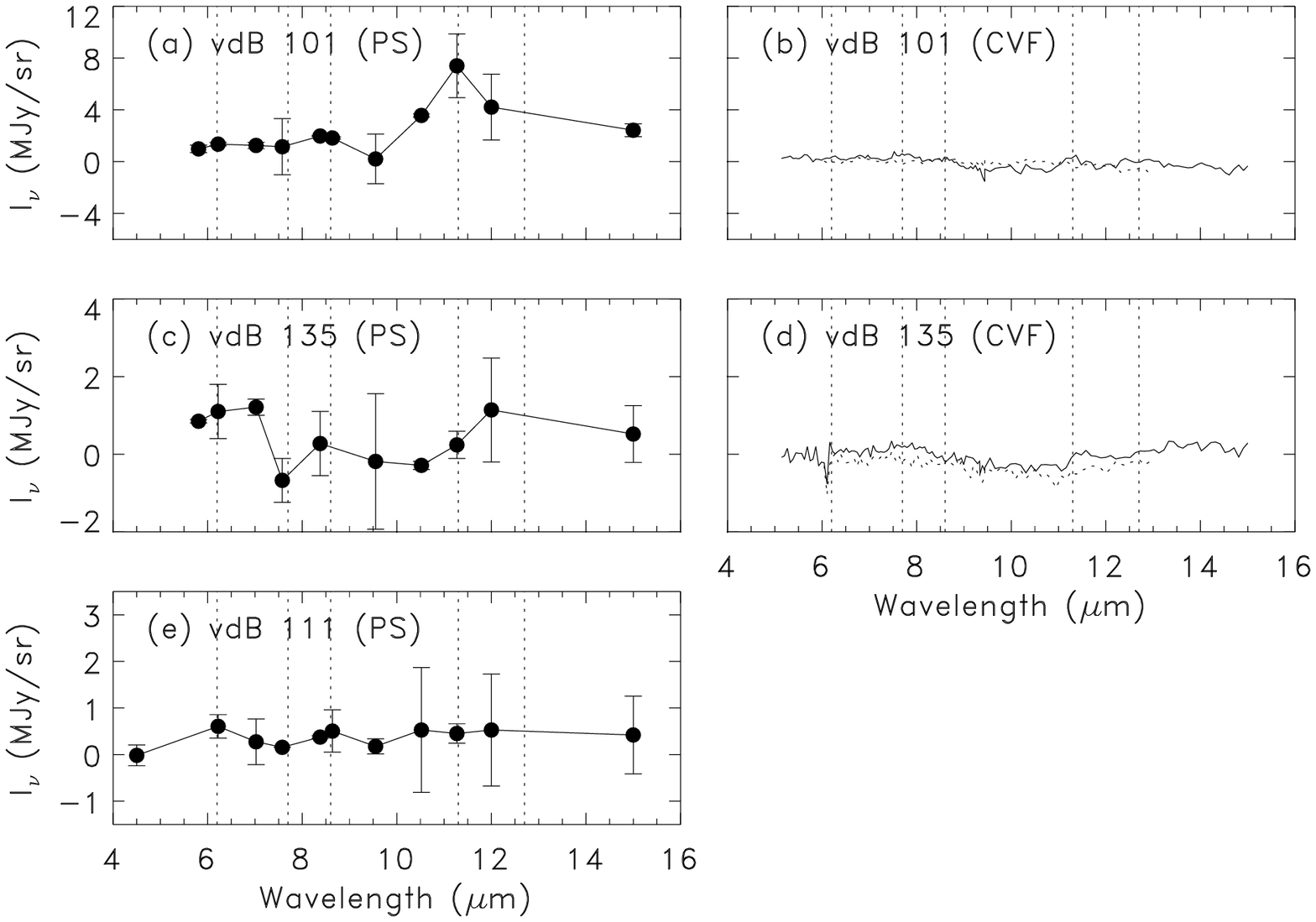}
\figcaption[]{Flux-rectified and sky-subtracted 5.1 -- 15 \micron\
spectra of ({\it a, b}) vdB\,101, ({\it c, d}) vdB\,135, and ({\it e})
vdB\,111, in $I_{\nu}$ (MJy sr$^{-1}$) vs.  $\lambda (\mu$m).
Wavelengths of the 6.2, 7.7, 8.6, 11.3, and 12.7 \micron\ IEFs are
marked ({\it dotted vertical lines}).  ({\it a, c, e}) Spectra from
position-switched observations at chosen CVF wavelengths ({\it filled
circles}).  Error bars show the difference between the cycle 1 and
cycle 2 source$-$sky rectified data at each wavelength.  ({\it b, d})
Spectra from complete CVF scan ({\it solid line}).  The difference
between the rectified increasing and decreasing wavelength scans of
the source-sky spectra ({\it dotted line}) is plotted as an estimate
of the uncertainty.  \label{iso3} }
\end{figure}

Figures \ref{iso1}, \ref{iso2}, and \ref{iso3} show our rectified,
sky-subtracted spectra of reflection nebulae from position-switched
observations and CVF scan observations.  Each reflection nebula
spectrum is a spatial {\it average} over a subsection of the image
containing the localized IEF emission.  The specifics of the spatial
subregions used to produce the spectra are given in Table
\ref{tab:intreg}.  For the position-switched data, where the image
contained the illuminating star, the star and the immediately adjacent
region containing a spurious arc-like reflection feature from the star
(typically pixels X,Y = [6--20, 11--24] of the array) were excluded
from the spatial integration.

The position-switched observations were the first performed in this
study.  The 11 unique wavelength measurements were chosen to measure
the strengths of the anticipated Class A interstellar IEFs (6.2, 7.6,
8.62, 11.22 $\micron$), their surrounding continua (4.5, 5.8, 9.5,
10.5, and 15.0 $\micron$) and broad spectral structures such as the 6
-- 9 $\micron$ bump and the 11 -- 13 $\micron$ plateau (7.0, 8.4, and
12.0 $\mu$m).

The CVF scan observations, which provide a complete CVF spectrum over
5.14 -- 15.1 $\mu$m, were performed second.  In nearly all the CVF
scans, there is a strong curvature at the start of the upward and
downward scans because of the detector's flux memory and the satellite
having slewed across the bright central star.  With such large flux
deviations, especially with the case of vdB\,17 and vdB\,59, the
response at the start of each scan could not entirely be corrected by
the aforementioned rectification procedures (\S3.1).  To avoid these
transient artifacts, we discarded data shortward of 8.0 $\mu$m for the
upward scan and data longward of 12.0 $\mu$m for the downward scan.
Our final sky-subtracted spectra are thus constructed from the
downward scan at 5.1 -- 8.0 $\mu$m, from an average of the upward and
downward scans at 8.0 -- 12.0 $\mu$m, and from the upward scan at 12.0
-- 15.1 $\mu$m.  The difference between the up and down scans is
plotted over a slightly wider range of wavelengths for each
CVF-scanned spectrum, to estimate the uncertainty in our CVF scan
spectra.

The 9 -- 10 $\mu$m segment of the CVF scan spectra also contain a
prominent artifact resulting from the steeply declining transmission
functions of the two CVF filter wheels near their wavelength overlap
region:  the long wavelength extreme of the CVF1 wheel (5.14 -- 9.44
\micron) and the short wavelength extreme of the CVF2 wheel (9.33 --
16.52 \micron).  Thus the 9 -- 10 $\mu$m spectral region, while shown
in some spectra, has been excluded from detailed analysis.

The position-switched spectrum of vdB\,133 (Fig \ref{iso1}{\it a})
displays the distinct signature of Class A interstellar IEFs at 6.2,
7.6, 8.62, and 11.22 $\mu$m.  The overwhelming majority of IEF spectra
observed fall within this class (Geballe 1997; Tokunaga 1997).  The
full CVF-scan spectrum (Fig.  \ref{iso1}{\it b}, and Fig.  1 of Paper
I) taken during the supplemental observing period strongly confirms,
both in location and relative intensities, the IEFs found in the more
coarsely sampled positioned-switched spectrum.

Figure \ref{iso2} displays the CVF-scanned spectra toward the
reflection nebulae vdB\,17 and vdB\,59.  The spectra of both nebulae
clearly display the classic 6.2, 7.7, 8.6, 11.3, and 12.7 \micron\
IEFs that characterize Class A type interstellar IEF spectra.  Both
these nebulae are illuminated by hot, UV-rich stars ({$T_{\rm eff}$} =
12,000 K and 19,000 K for vdB\,17 and vdB\,59, respectively) and are
typical of those that display strong interstellar IEFs.

Figure \ref{iso3} presents 5 -- 15 $\mu$m spectra of vdB\,101,
vdB\,111, and vdB\,135 obtained with ISOCAM.  Both the
position-switched spectra and the fully-sampled CVF spectra of each
source are shown, with the exception of vdB\,111, toward which only
position-switched observations were made.  The fully sampled spectra
of vdB\,101 and vdB\,135 are generally consistent with their
position-switched counterparts obtained toward slightly different
positions (Table \ref{tab:intreg}).  The source and background levels
of vdB\,101, vdB\,111 and vdB\,135 measured by ISOCAM at 12 $\micron$\
are consistent with those of {\it IRAS}.  The reflection nebulae
vdB\,101, vdB\,111, and vdB\,135 are all illuminated by cool, UV-poor
stars ({$T_{\rm eff}$} = 5,000 K, 7,300 K, and 3,600 K, respectively).
These three nebulae have very low 12 $\micron$ {\it IRAS} fluxes,
ranging between 0.3 and 0.8 MJy sr$^{-1}$.  No IEFs, or continuum
emission, are apparent in any of the source\,$-$\,sky spectra above
the noise level of the observations, $\sim$1 MJy sr$^{-1}$.

The task of detecting IEFs toward vdB\,101, vdB\,111, and vdB\,135,
even with ISOCAM operating under optimal conditions, was understood to
be a difficult one.  The predicted sensitivity of ISOCAM ($\sim$0.19
MJy sr$^{-1}$ with 200\,s integration time per filter in the
position-switched observations) should have been sufficient, in
principle.  In practice, our actual ISOCAM sensitivity limit was
$\sim$1 MJy sr$^{-1}$ for our observations, due to the faintness of
the three nebulae compared to the relatively high background emission
from diffuse Galactic emission along the same line of sight.  The
larger than anticipated uncertainty is attributed primarily to the
limitations of the rectification procedure to completely correct for
the detector effects and the uncertainties in the dark-current
subtraction procedure required for the rectification process
(discussed in \S3).  To achieve the expected sensitivity of 0.19 MJy
sr$^{-1}$ on top of a typical background level of 20 MJy sr$^{-1}$
would require that our flux rectification procedure be accurate to
$\sim$1\%.  Comparison of the upwards and downwards CVF scans
indicated that in practice the rectification is only accurate to
$\sim$5\%.

\section{Discussion}

In the following, we present a quantitative analysis of the IEFs as a
function of the UV radiation field.  We explore the effect of both the
hardness of the UV field ($T_{\rm eff}$ of the illuminating star) and
the absolute intensity of the UV field ($G_0$).  We expand our
analysis over a wider range of $T_{\rm eff}$ and $G_0$ by comparing
our observed sample to published spectra of vdB\,106 ($\rho$ Oph;
Boulanger et al.  1996), and vdB\,139 (NGC\,7023; D.  Cesarsky et al.
1996).  We also examine spectra at different distances from the
illuminating star, for vdB 17 and vdB 59.  This examines the effect of
constant $T_{\rm eff}$ but varying $G_0$.

\subsection{Quantifying the UV field toward reflection nebulae}

The motivation for our {\it ISO} observations of reflection nebulae is
the observed lack of dependence on $T_{\rm eff}$ of $\Delta \nu
I_\nu$(12 $\mu$m)/$I_{\rm bol}$(FIR) in reflection nebulae (Sellgren
et al.  1990; see \S1 and Table \ref{tab:stelpar}).  This is puzzling
because current dust models predict that the {\it IRAS} 12 $\mu$m band
is dominated by PAH emission and that PAHs contribute more to the
total interstellar extinction curve at UV than visible wavelengths
(D\'esert et al.  1990; Joblin et al.  1992; Siebenmorgen \& Kr\"ugel
1992; Schutte et al.  1993; Dwek et al.  1997; Li \& Greenberg 1997;
Silva et al.  1998).  Because we are interested in analyzing {\it ISO}
spectra as a function of UV strength, we need to quantify the UV
strength in our sources.  We have parameterized the UV field in four
ways.

Our first method of characterizing the UV field is simply by $T_{\rm
eff}$ for the illuminating star of each reflection nebula (Table
\ref{tab:stelpar}).  Hotter stars emit a larger fraction of their
total energy in the UV.  This is complicated, however, by the
difference in $T_{\rm eff}$ for the two stars which illuminate vdB 133
($T_{\rm eff}$ = 6,800 K and 12,000 K; the cooler star is more
luminous).

Another measure of the UV field is the ratio of the stellar flux
emitted at UV wavelengths to the total stellar flux.  This simplifies
the treatment of the UV field in vdB 133, since this ratio equals the
sum of the UV fluxes of the two stars divided by the sum of the total
flux of the two stars.  The relationship between $T_{\rm eff}$ and the
ratio of UV to total stellar flux is, however, defined by the longest
wavelength included in the range of UV wavelengths.  We have explored
two ranges of UV wavelengths, both inspired by previous theoretical,
observational, and laboratory work on PAHs and the IEFs.

The low ratio of visible to UV absorption, for the distribution of PAH
sizes included in published models of interstellar absorption and
emission, occurs because large molecules in general do not have
electronic transitions beyond a cutoff wavelength, $\lambda _ {\rm
cutoff}$, proportional to the molecule's length (Platt 1956).  Values
of $\lambda _ {\rm cutoff}$ = 300 -- 500 nm have been estimated for
neutral PAHs containing roughly 40 carbon atoms (Crawford, Tielens, \&
Allamandola 1985; Allamandola et al.  1989; L\'eger et al.  1989;
D\'esert et al.  1990; Joblin et al.  1992; Schutte et al.  1993).
Ionized PAHs absorb at longer wavelengths than their neutral
counterparts (Salama et al.  1996 and references therein).
Approximate formulae for the relation between $\lambda _ {\rm cutoff}$
and PAH size have been published for neutral PAHs (D\'esert et al.
1990; Schutte et al.  1993) and for ionized PAHs (D\'esert et al.
1990; Salama et al.  1996).

Thus, as our second method of parameterizing the UV field, we define a
UV wavelength range of 91 -- 400 nm.  In this case, the fraction of
total stellar flux emitted at UV wavelengths is {$F_{(\lambda < 400
{\rm nm})}$/$F_{total}$}.  We chose $\lambda _ {\rm cutoff}$ = 400 nm
for this definition because it corresponds to the value of $\lambda _
{\rm cutoff}$, averaged over a distribution of PAH sizes and charge,
at which current dust models predict the contribution of PAHs to the
total interstellar extinction curve becomes negligible (D\'esert et
al.  1990; Joblin et al.  1992; Siebenmorgen \& Kr\"ugel 1992; Dwek et
al.  1997; Silva et al.  1998).  Note 400 nm falls just longward of
the observational dividing line between UV and visible wavelengths.

Our third method of parameterizing the UV field is to define a UV
wavelength range of 91 -- 240 nm.  Many previously published
theoretical and observational papers on the ISM characterize the
ambient radiation field by $G_0$, the strength of the 91 -- 240 nm UV
radiation field.  Examples of the prevalence of $G_0$ as a measure of
the UV energy density include observational and theoretical studies of
the 12 $\mu$m {\it IRAS} emission in our own and other galaxies
(Draine \& Anderson 1985; Ryter, Puget, \& P\'erault 1987; Boulanger
et al.  1988; D\'esert et al.  1990; Helou, Ryter, \& Soifer 1991;
Rowan-Robinson 1992), analysis of the observed strengths of the IEFs
from recent {\it ISO} and {\it IRTS} spectra (Boulanger et al.  1996,
1998a, 1998b; Onaka 1999), and models exploring the ionization state
of PAHs and their effect on the heating and chemistry of different
regions of the ISM (Allamandola et al.  1989; Bakes \& Tielens 1994,
1998; Spaans et al.  1994; Wolfire et al.  1995; Salama et al.  1996;
Dartois \& d'Hendecourt 1997).  For a UV range of 91 -- 240 nm, the
fraction of total stellar flux emitted at UV wavelengths is
{$F_{(\lambda < 240 {\rm nm})}$/$F_{total}$}.

\begin{figure}
\figurenum{5}
\plotone{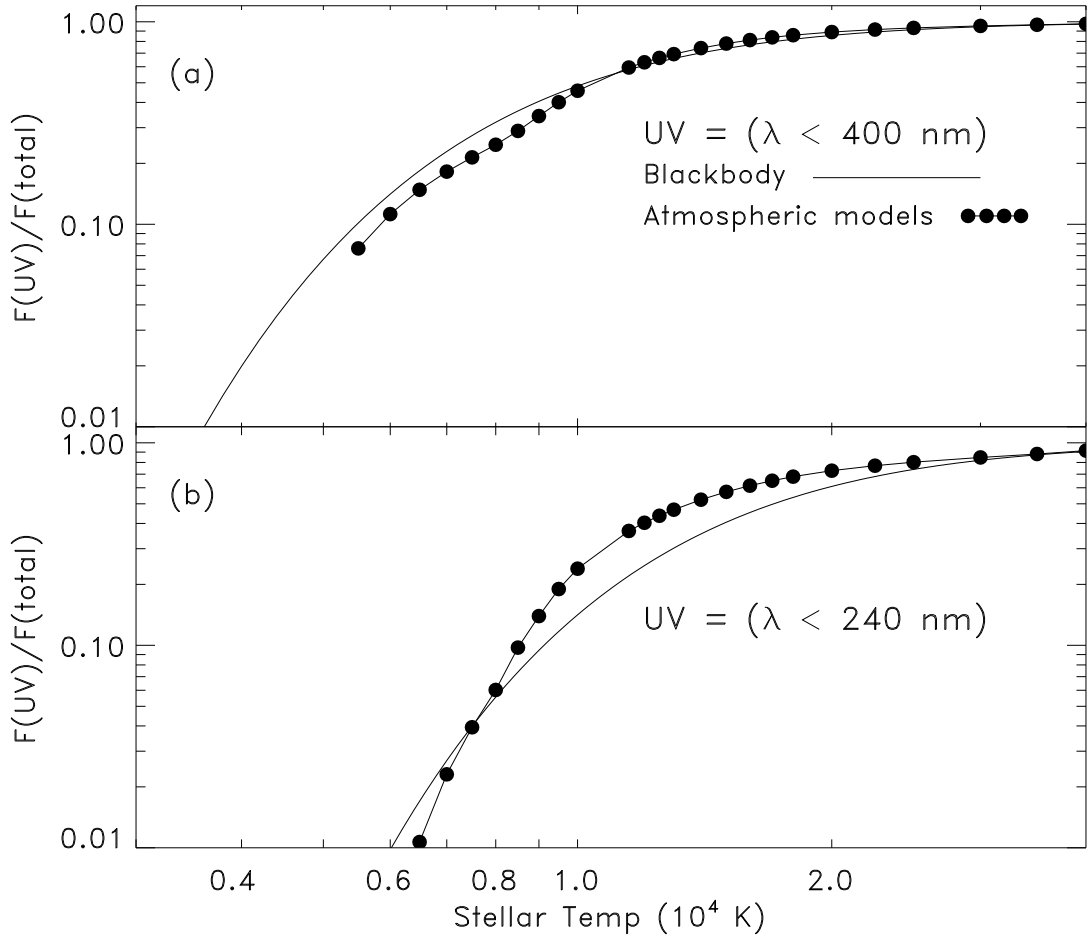}
\figcaption[]{Plots of the fraction of UV emission to total emission
as a function of stellar temperature, calculated from ({\it connected
dots}) Kurucz (1979) atmospheric models with log($g$) = 4.0 and from
({\it solid line}) a blackbody.  The UV range is defined as ({\it a})
$\lambda$\ $<$ 400 nm or ({\it b}) $\lambda$\ $<$ 240 nm. \label{uvrat} }
\end{figure}

Figure \ref{uvrat} displays {$F_{(\lambda < 400 {\rm
nm})}$/$F_{total}$}\ and {$F_{(\lambda < 240 {\rm nm})}$/$F_{total}$}\
as a function of $T_{\rm eff}$, determined using two different stellar
models.  We use Kurucz (1979) stellar atmospheric models for $T_{\rm
eff}$ $>$ 5,500\,K, with log($g$) = 4.0.  We find, however, that the
results are not sensitive to the value of log($g$) adopted.  Since the
Kurucz models do not extend to cooler temperatures, we use a blackbody
(Planck) function to model stars with $T_{\rm eff}$ $<$ 5,500\,K.
Figure \ref{uvrat} shows that a blackbody slightly overestimates
{$F_{(\lambda < 400 {\rm nm})}$/$F_{total}$}\ and {$F_{(\lambda < 240
{\rm nm})}$/$F_{total}$}\ for the coolest stars for which we have
Kurucz models.  Figure \ref{uvrat} and Table \ref{tab:stelpar} show
that {$F_{(\lambda < 400 {\rm nm})}$/$F_{total}$}\ drops by a factor
of 10, and {$F_{(\lambda < 240 {\rm nm})}$/$F_{total}$}\ falls by over
a factor of 100, as $T_{\rm eff}$ decreases from 21,000\,K to
5,000\,K.

Our fourth method of parameterizing the UV field incident on a
particular nebular position is by $G_0$ (sometimes called $\chi$ or
$u$), the strength of the 91 -- 240 nm UV radiation field in units of
the local interstellar radiation field.  A value of $G_{0}$ = 1
corresponds to a UV radiation field of 1.6 $\times$\ 10$^{-3}$ erg
s$^{-1}$ cm$^{-2}$ at $\lambda$ = 91 -- 240 nm (Habing 1968).  Table
\ref{tab:gfac} presents the distance to each source and the projected
angular separation between the illuminating stars and the observed
nebular region used to derive $G_{0}$.  A single value of $G_0$
suffices for vdB 106, vdB 133, and vdB 139, where the observed
spectrum was obtained over a narrow range of distances from the
central star, and for vdB 101, vdB 111, and vdB 135, which we did not
detect with {\it ISO} spectroscopy.  For vdB 17 and vdB 59, we were
able to obtain spectra at a range of values of stellar distance and
thus $G_0$.  Table \ref{tab:gfac} gives the value of $G_0$ at a
typical nebular position for these two nebulae; we discuss how the
spectra of vdB 17 and vdB 59 depend on nebular position (and thus
$G_0$) in a later section.

We summarize the UV field for our observed sources and published
comparison sources, as measured by $T_{\rm eff}$, {$F_{(\lambda < 400
{\rm nm})}$/$F_{total}$}, {$F_{(\lambda < 240 {\rm
nm})}$/$F_{total}$}, and $G_0$, in Tables \ref{tab:stelpar} and
\ref{tab:gfac}.  The observed values of $\Delta \nu I_\nu$(12
$\mu$m)/$I_{\rm bol}$(FIR) (Sellgren et al.  1990; see \S1 and Table
\ref{tab:stelpar}) are independent of {\it all} these measures of the
UV strength.

\subsection{PAH ionization state in reflection nebulae}

Another reason to quantify the UV field in reflection nebulae is to
search for spectroscopic differences between neutral and ionized PAHs.
Interstellar PAHs are predicted to be mainly positively charged and
neutral in reflection nebulae illuminated by B stars, and mainly
neutral and negatively charged in the diffuse interstellar medium
(Bakes \& Tielens 1994, 1998; Salama et al.  1996; Dartois and
d'Hendecourt 1997).  Predicted differences between neutral and ionized
PAHs include changes both in the central wavelengths and in the
relative strengths of PAH features (De Frees et al.  1993; Szczepanski
\& Vala 1993; Hudgins, Sandford, \& Allamandola 1994; Pauzat, Talbi,
\& Ellinger 1995, 1997; Langhoff 1996; Cook \& Saykally 1998;
Allamandola, Hudgins, \& Sandford 1999; Hudgins \& Allamandola 1999b).

The first ionization potential of PAHs lies in the range 4 -- 8 eV,
with a tendency for the ionization potential to decrease with
increasing PAH size (Bakes \& Tielens 1994; Salama et al.  1996).
Hudgins \& Allamandola (1999a) conclude from the wavelength spacing of
the 6.2 and 7.7 $\mu$m IEFs observed in the ISM that these IEFs arise
from ionized PAHs with 50 -- 80 carbon atoms.  A PAH with 50 -- 80
carbon atoms has a first ionization potential of 5.8 -- 6.2 eV,
corresponding to an ionization wavelength of 200 -- 210 nm (Bakes \&
Tielens 1994).  The value of $G_0$ (with $\lambda _ {\rm cutoff}$ =
240 nm) is thus a rough indicator of the intensity of stellar
radiation capable of ionizing PAHs at any given nebular position.

Theory predicts that the mean ionization state of interstellar PAHs is
a function of $G_0$ (Bakes \& Tielens 1994, 1998; Salama et al.  1996;
Dartois and d'Hendecourt 1997).  If the spectral features we observe
in reflection nebulae are due to PAHs, then we would expect to detect
changes both in the central wavelengths and in the relative strengths
of IEFs between nebulae with different values of $G_0$.  One goal of
our observations is to test this prediction.

\subsection{Spatially averaged reflection nebulae spectra}

Localized enhancements of IEF emission, at 6.2, 7.7, 8.6, 11.3, and
12.7 $\mu$m, were detected in the source\,$-$\,sky spectra in three of
the six nebulae observed:  vdB\,17, vdB\,59 and vdB\,133.  These three
nebulae are illuminated by the hotter stars in our sample, with
$T_{\rm eff}$ = 6,800 K -- 19,000 K.  Our $\sim$1 MJy sr$^{-1}$ upper
limits on the IEF emission in vdB\,101, vdB\,111, and vdB\,135 are
consistent with their 12 $\micron$ {\it IRAS} surface brightnesses of
0.3 -- 0.8 MJy sr$^{-1}$ (Sellgren et al.  1990).  These faint nebulae
have cooler illuminating stars, with $T_{\rm eff}$ = 3,600 K -- 7,300
K.

\begin{figure}
\figurenum{6}
\plotone{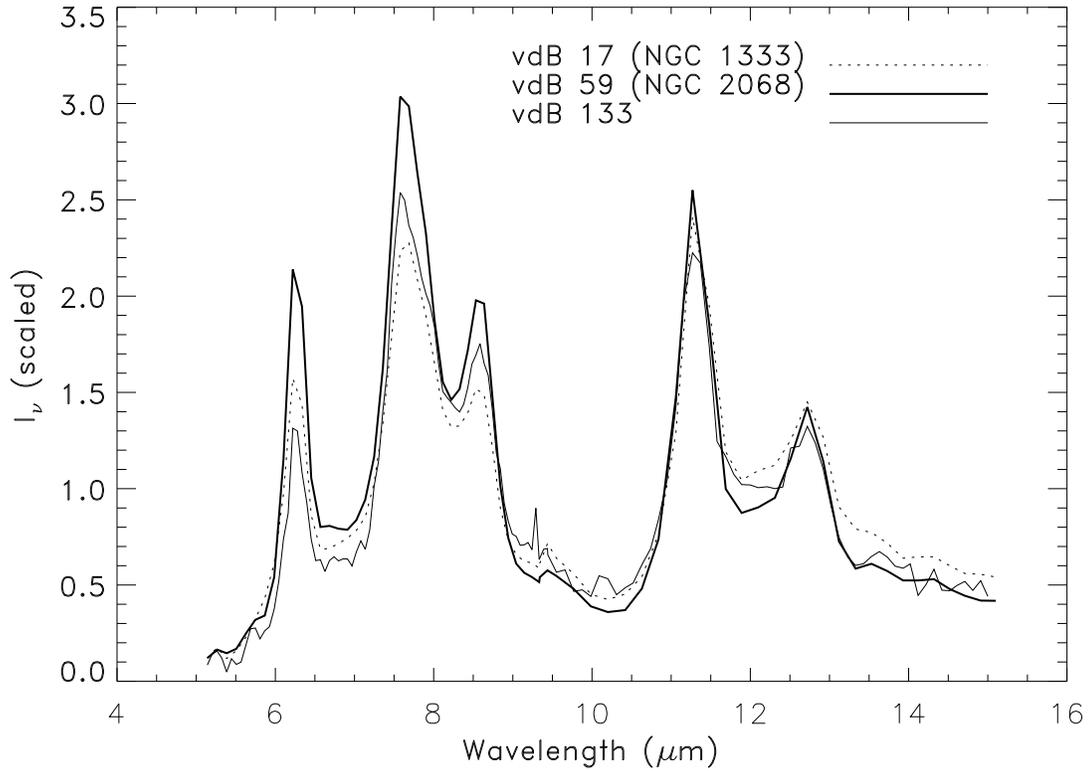}
\figcaption[]{Superposed spectra of ({\it dotted line}) vdB\,17 (NGC
1333), ({\it thick solid line}) vdB\,59 (NGC 2068), and ({\it thin
solid line}) vdB\,133.  Spectra are $I_{\nu}$ (MJy sr$^{-1}$), divided
by their intensity in the {\it IRAS} 12 $\mu$m broad-band filter (MJy
sr$^{-1}$), vs.  $\lambda$($\mu$m).  The fraction of total stellar
flux emitted at UV wavelengths, for the illuminating stars of these
reflection nebulae, is lowest for vdB\,133 and highest for vdB\,59.
\label{lcomp} }
\end{figure}

Figure \ref{lcomp} displays the superposed spectra of vdB\,17,
vdB\,59, and vdB\,133, those nebulae we observe having IEF emission in
excess of the adjacent sky emission.  These spectra (from Figs.
\ref{iso1} and \ref{iso2}) are spatial averages over the emission
region in each reflection nebula, defined by visual inspection of the
images.  Each spectrum is scaled by its average intensity over the 12
$\mu$m {\it IRAS} band ($\lambda\ \simeq\ 7.5 - 15\ \mu$m).  We choose
this scaling to illustrate the IEF peak and continuum levels (in MJy
sr$^{-1}$), normalized to an {\it IRAS} 12 $\mu$m intensity of 1 MJy
sr$^{-1}$, for typical reflection nebulae.  Such illustration is
useful for planning for future mid-infrared spectroscopy with {\it
SIRTF} and {\it SOFIA} of sources detected by the {\it IRAS} all-sky
survey.

The reflection nebulae spectra in Figure \ref{lcomp}, including their
continua, broad emission bumps, and narrow IEFs, are roughly similar
in shape despite the widely varying range of conditions quantified in
Tables \ref{tab:stelpar} and \ref{tab:gfac}.  The normalized peak
fluxes of the different IEFs vary by less than a factor of two from
source to source.  More importantly, these variations show no
systematic dependence on $T_{\rm eff}$.  No evidence is seen among
these three spectra for disappearance of IEFs, appearance of new
spectral features, shifts in IEF wavelengths, or dramatic changes in
the strengths of IEFs relative to the broad emission bumps or the
continuum emission.

\subsubsection{Quantifying the IEF Features}

\begin{figure}
\figurenum{7}
\plotone{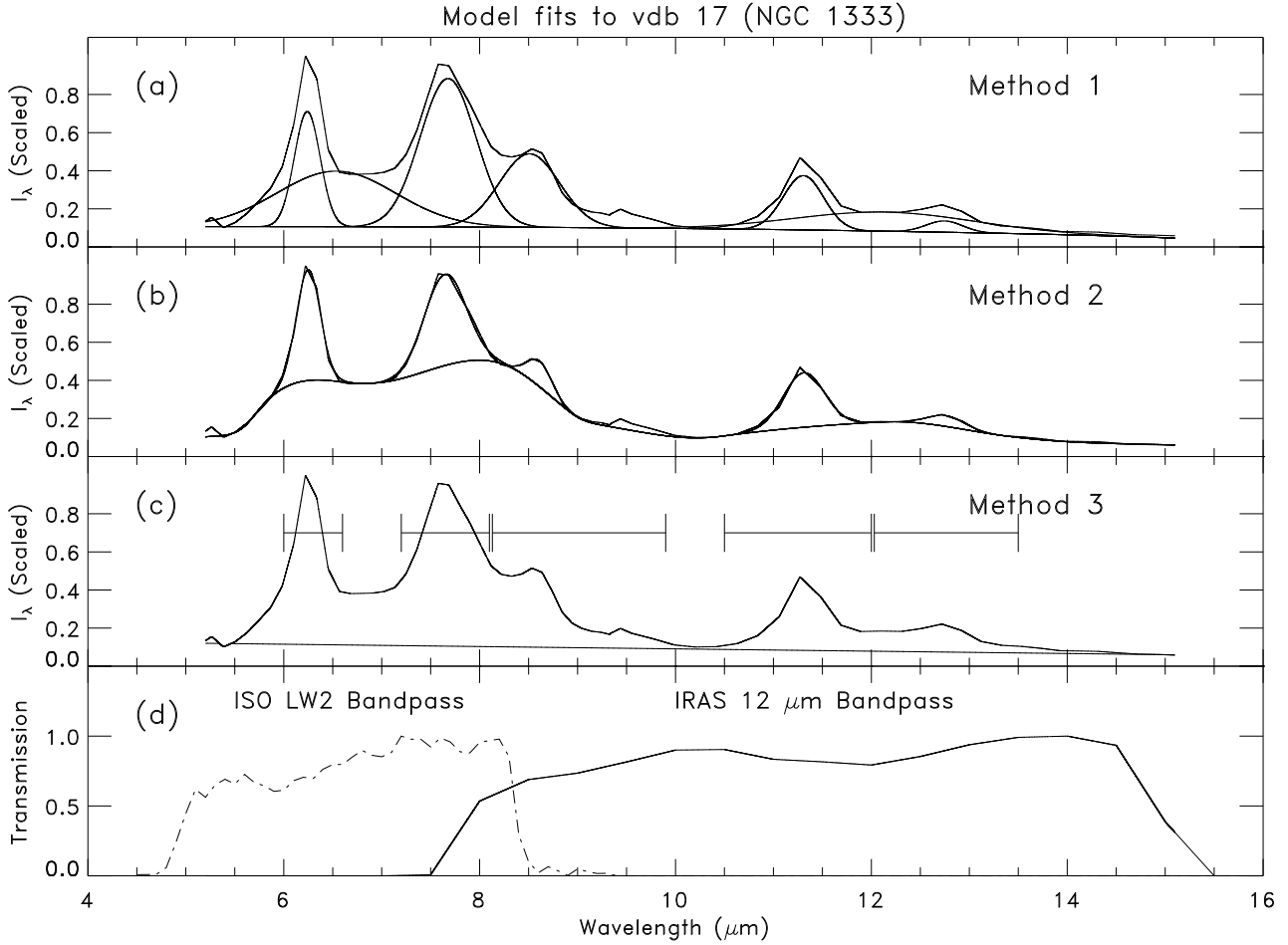}
\figcaption[]{The 5 -- 15 \micron\ spectrum of vdB\,17 (NGC\,1333)
superposed with feature and continuum components fitted by ({\it a})
Method 1, ({\it b}) Method 2, and ({\it c}) Method 3 (see text).  The
integration regions adopted for the various IEFs in Method 3 ({\it c})
are marked ({\it horizontal bars}).  The ISO LW\,2 and IRAS 12 $\mu$m
filter transmission functions are also plotted ({\it d}).  Note
spectra are plotted as $I_{\lambda}$(erg cm$^{-2}$ s$^{-1}$ sr$^{-1})$
vs.  $\lambda (\mu$m), which differs from other figures.  \label{lfit}
}
\end{figure}

The quantitative fluxes of the narrow IEFs, broad emission bumps, and
continuum are quite sensitive to how the spectrum is divided into
separate spectral components.  Figure \ref{lfit} displays the spectrum
of vdB\,17, to illustrate the range of possibilities for dividing the
emission between the different spectral components.

Figure \ref{lfit}{\it a} shows Method 1, in which we simultaneously
fit Gaussians to the five {\it narrow} IEFs (at 6.2, 7.7, 8.6, 11.3
and 12.7 \micron) and to two {\it broad} emission features at $\sim$7
and 12 $\micron$.  The latter two features represent the 6 -- 9 $\mu$m
``bump'' (Cohen et al.  1986; Bregman et al.  1989) and the 11 -- 13
$\mu$m ``plateau'' (Cohen, Tielens, \& Allamandola 1985).  The {\it
continuum} is approximated by a spline fit through the data points at
$\sim$5.4, 10.3, and 14.8 $\micron$.  The central wavelengths, widths,
and heights of the Gaussians were all free parameters in the fit.
Boulanger et al.  (1998a) have suggested that the IEFs are best fit by
Lorentzians.  We used Gaussians rather than Lorentzians for the IEFs
because they provide a better spectral fit to the IEFs observed in NGC
7023 at $R$ = 1000 (Sellgren et al.  1999).

Figure \ref{lfit}{\it b} shows Method 2, in which we simultaneously
fit Gaussians to the five narrow IEFs (at 6.2, 7.7, 8.6, 11.3 and 12.7
\micron).  The continuum is modeled by a spline fit to all spectral
data points outside that of the narrow IEFs; we assume the broad
emission features at 6 -- 9 $\mu$m and at 11 -- 13 $\mu$m are not
distinct from the continuum but rather are simply undulations in the
continuum level in this case.  Again the central wavelengths, widths,
and heights of the Gaussians are free parameters in the fit.

Figure \ref{lfit}{\it c} shows Method 3, in which we modeled the
continuum by a linear fit to the spectral data points at $\sim$5.3 and
15.0 $\mu$m, the wavelength extremes of the observations.  No broad
emission features are assumed in this case.  The fluxes of the five
narrow IEFs were found by directly integrating over the
continuum-subtracted spectra, the method most used in previous
studies.  The integration limits adopted for the 6.2, 7.7, 8.6, 11.3
and 12.7 \micron\ lines were (6.0 -- 6.6 $\mu$m), (7.2 -- 8.1 $\mu$m),
(8.1 -- 9.9 $\mu$m), (10.5 -- 12.0 $\mu$m), (12.0 -- 13.5 $\mu$m),
respectively, as indicated in Figure \ref{lfit}{\it c}.  The limits of
the 6.2, 7.7, 8.6, and 11.3 $\micron$ IEFs are those used by Lu
(1998).  For the 12.7 \micron\ IEF, which Lu does not include in his
analysis, we integrate over a wavelength range similar in extent
($\Delta\lambda$ = 1.5 $\mu$m) to that used for the 11.3 \micron\
line.

Figure \ref{lfit} illustrates the sensitivity of the estimated
strengths of IEFs, broad features, and continuum to whether Method 1,
2, or 3 is used to fit the spectrum.  The Gaussian used to fit the 6
-- 9 $\mu$m broad bump in Method 1 is centered near 6.7 $\mu$m, and as
a result the narrow 7.7 $\mu$m IEF and especially the 8.6 $\mu$m IEF
appear stronger.  The higher adopted continuum level at 6 -- 9 $\mu$m
in Method 2 is centered closer to 7.5 $\mu$m, which results in a
somewhat weaker 7.7 $\mu$m IEF and a much weaker 8.6 $\mu$m IEF.  The
choice of a linear continuum (fitted to points at 5.3 and 15.0 $\mu$m)
in Method 3 consistently produce the highest 6 -- 9 $\mu$m feature
strengths of among the methods.  The choice of broad integration
limits for the 8.6 $\mu$m feature (8.1 -- 9.9 $\mu$m) result in
strengths comparable to or exceeding that of the 6.2 $\mu$m feature.
Method 2 and Method 3 represent two extremes for the treatment of the
6 -- 9 $\mu$m region.

Previous studies of the relative strengths of different IEFs vary in
their choices of continuum placement and integration limits in
wavelength; the choice of continuum by Cohen et al.  (1986, 1989) and
Lu (1998) in their analyses is similar to Method 2, while that of
Jourdain de Muizon, d'Hendecourt, \& Geballe (1990) and Zavagno, Cox,
\& Baluteau (1992) is more like that of Methods 1 and 3.  The
sensitivity of our feature strengths to continuum placement,
especially, demonstrates how much care must be exercised in
intercomparing the results of different studies.

\subsubsection{Comparison of Nebulae} 

Table \ref{tab:linerat} presents our quantitative results for the
strengths of the IEFs, broad emission features, and continuum in
vdB\,17, vdB\,59, vdB\,133 (Paper I), vdB\,139 (NGC\,7023; D.
Cesarsky et al.  1996), and vdB\,106 ($\rho$ Oph; Boulanger et al.
1996) derived by three different methods for defining the IEFs and
continuum.  The methods by which the continuum and IEFs are defined,
and thus the feature ratios quoted, vary between studies; the three
methods employed here are representative of those typically used.  The
individual feature strengths are given in units of integrated line
flux (ergs s$^{-1}$ cm$^{-2}$ sr$^{-1}$) divided by the total
integrated 11.3 $\mu$m line flux (in ergs s$^{-1}$ cm$^{-2}$
sr$^{-1}$).

\begin{figure}
\figurenum{8}
\plotone{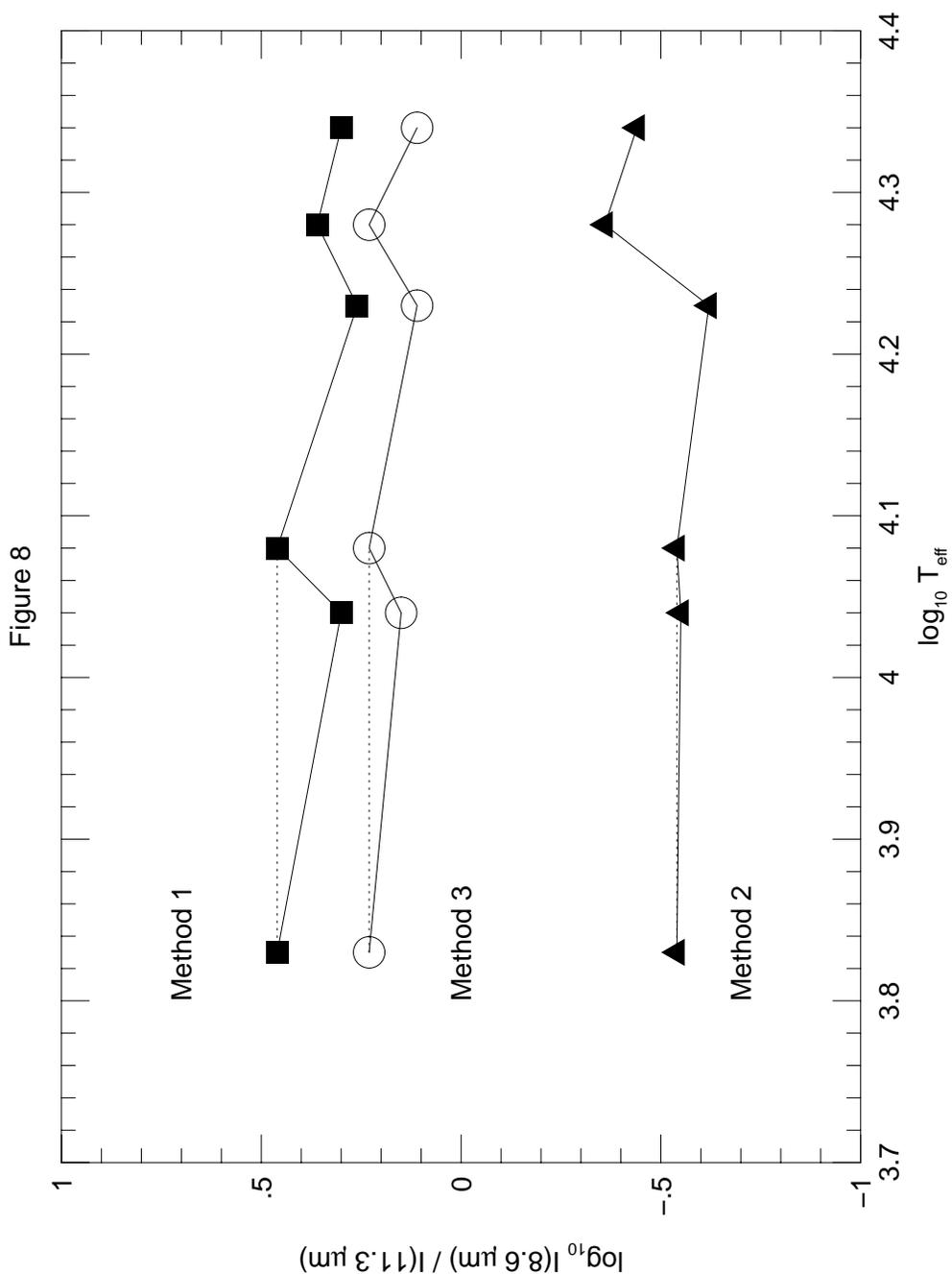}
\figcaption[]{Plot displaying log[$F$(8.6 $\mu$m)/$F$(11.3 $\mu$m)],
the ratio of integrated 8.6 $\mu$m and 11.3 $\mu$m IEF fluxes (erg
cm$^{-2}$ s$^{-1}$ sr$^{-1}$), vs.  log($T_{\rm eff}$), the effective
temperature of the illuminating star(s) of each reflection nebula.
Values of log[$F$(8.6 $\mu$m)/$F$(11.3 $\mu$m)] determined by Method 1
({\it filled squares}), Method 2 ({\it filled triangles}), and Method
3 ({\it open circles}) are compared (see text).  Values of log[$F$(8.6
$\mu$m)/$F$(11.3 $\mu$m)] for each Method are connected ({\it solid
lines}) for clarity.  Values of log[$F$(8.6 $\mu$m)/$F$(11.3 $\mu$m)]
are plotted at two values of $T_{\rm eff}$, corresponding to the
hotter and cooler illuminating star of vdB 133 (points connected by
{\it dotted lines}).  \label{lrat} }
\end{figure}

Table \ref{tab:linerat} shows that the relative strengths we derive
for the IEFs, broad emission features, and continuum emission are very
similar for all the sources we analyze, within a given method for
decomposing the spectrum.  The relative IEF strengths for any given
source, however, are very sensitive to the method chosen.  Figure
\ref{lrat} illustrates this with the 8.6 to 11.3 $\mu$m IEF ratio,
$I$(8.6 $\mu$m)/$I$(11.3 $\mu$m), plotted against $T_{\rm eff}$.  We
originally chose this ratio because it is predicted to be much
stronger for ionized PAHs than for neutral PAHs (De Frees et al.
1993; Szczepanski \& Vala 1993; Hudgins et al.  1994; Pauzat et al.
1995, 1997; Langhoff 1996; Cook \& Saykally 1998; Allamandola et al.
1999; see \S5.4.1).  The value of $T_{\rm eff}$ (Table
\ref{tab:stelpar}) is a measure of the hardness of the UV radiation
field, which is independent of any adopted value of $\lambda_{\rm
cutoff}$, and which should be similar for all nebular positions
included in the spatially averaged spectrum of each source.  Figure
\ref{lrat} shows, however, that systematic differences in the mean
values of $I$(8.6 $\mu$m)/$I$(11.3 $\mu$m), measured by Methods 1, 2,
and 3, are much larger than the source-to-source scatter of $I$(8.6
$\mu$m)/$I$(11.3 $\mu$m) for any particular method.

\subsubsection{Convolution of the Spectra with Broad-band Filters}

Much effort has gone into characterizing the mid-infrared emission of
the ISM using broadband filters, which provide limited spectral
information but higher sensitivity than narrowband filters or
spectrometers.  The entire sky has been surveyed in the {\it IRAS} 12
$\mu$m filter, and many ISOCAM studies have been conducted with the
LW2 and LW3 filters, centered at 6.75 and 15 $\mu$m respectively.
Note that the ISOCAM LW10 filter and the ISOPHOT 11.5 $\mu$m filter
are designed to be the same as the {\it IRAS} 12 $\mu$m filter.  Each
of these broad filter bandpasses covers a mixture of IEFs, broad
emission features, and continuum emission from interstellar dust and
molecules.  Table \ref{tab:linerat} therefore presents the expected
{\it IRAS} 12 $\mu$m source fluxes and ISOCAM LW\,2 source fluxes
found by convolving the source spectra with the 12 $\mu$m {\it IRAS}
band transmission profile and LW\,2 transmission profile (Fig.
\ref{lfit}{\it d}), respectively.

We have also investigated what fraction of the flux within these
broadband filters comes from each spectral component in the reflection
nebulae we have observed.  Table \ref{tab:lineband} gives the relative
contributions of the IEFs, broad emission features, and continuum to
the broadband fluxes measured with broadband filters in common use.
We derived these results by convolving the ISOCAM CVF spectra with the
transmission profile of the LW2 (6.75 $\mu$m) filter on ISOCAM and the
{\it IRAS} 12 $\micron$\ filter.  Again, we find the estimated
percentage contributions of IEFs, broad features, and continuum to the
{\it IRAS} 12 $\mu$m filter and the ISOCAM LW2 filter are remarkably
consistent from nebula to nebula, within a given Method (1, 2, or 3),
but the absolute values for any nebula are quite sensitive to which
Method (1, 2, or 3) is used to fit the spectrum.  The choice of the
continua via Method 3 gives the maximal values of the narrow IEF/total
ratios, while those of Method 2 represent the lower extreme.  For the
{\it IRAS} 12 $\mu$m filter, the narrow IEFs contribute typically
$\sim$20\% (Method 1), $\sim$40\% (Method 2), and $\sim$60\% (Method
3) of the total flux depending on the method used to define the
continuum.  All methods imply that 40 -- 80\% of the {\it IRAS} 12
$\mu$m flux is due to continuum emission and/or broad emission
features, rather than the narrow IEFs.

\subsection{Spatial variations within individual reflection nebulae}

The spectra for vdB\,17 and vdB\,59 plotted in Figures \ref{iso2} and
\ref{lcomp}, and analyzed in Figures \ref{lfit} and \ref{lrat} and
Table \ref{tab:linerat}, have been integrated over a wide range of
projected distance from the central star.  In this next section we
explore how the spectral features within a single source image depend
on position for vdB 17 and vdB 59.

\begin{figure}
\figurenum{9}
\plotone{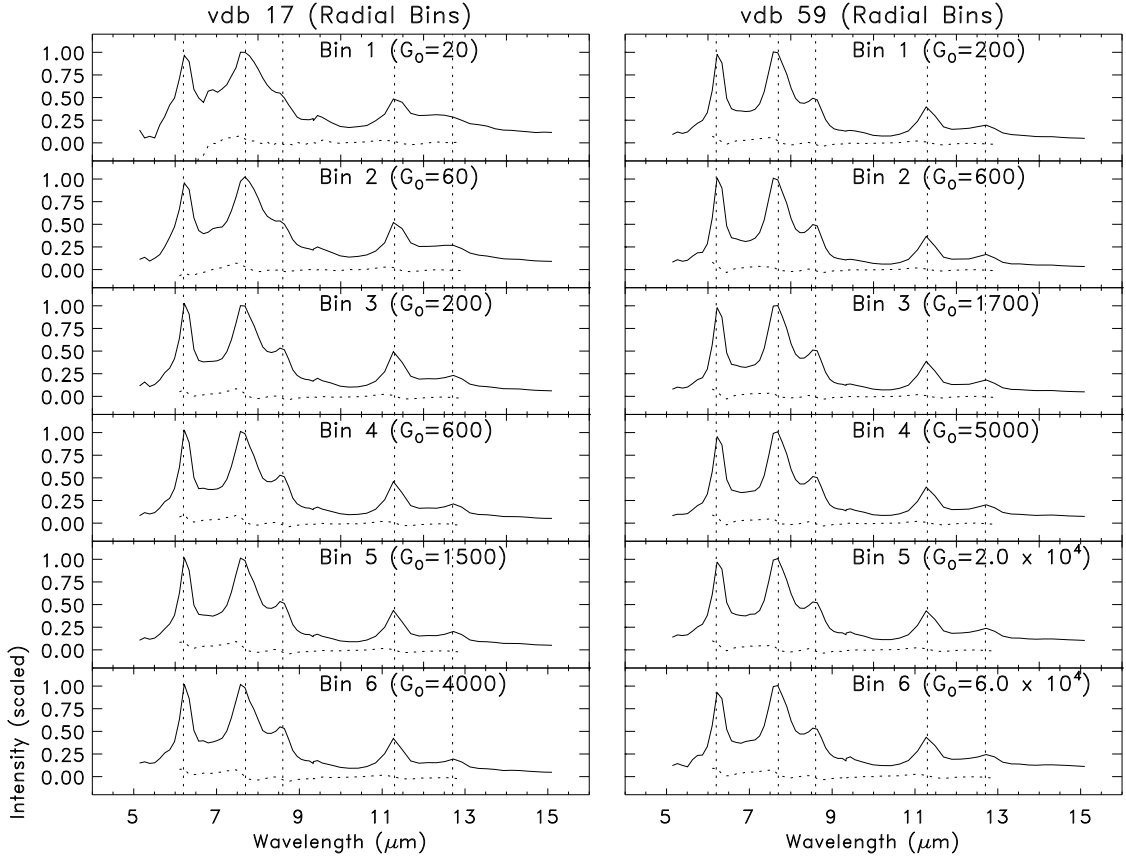}
\figcaption[]{ Spectra ({\it solid lines}) of vdB\,17 ({\it left}) and
vdB\,59 ({\it right}), averaged over radial bins of projected distance
from the illuminating star.  Radial bins are characterized by the UV
field, $G_0$, incident on each nebular annulus (see text).  $G_0$ is
in units of 1.6\,$\times$\,10$^{-3}$ ergs s$^{-1}$ cm$^{-2}$, the
local interstellar radiation field as determined by Habing (1968).
Bins with values of $G_0$ = 20, 60, 200, 600, 1500, and 4000 ({\it top
to bottom}) are shown for vdB\,17.  Bins with values of $G_0$ = 200,
600, 1700, 5000, 2.0 $\times$\ 10$^4$, and 6.0 $\times$\ 10$^4$ ({\it
top to bottom}) are shown for vdB\,59.  Spectra are $I_{\nu}$ (MJy
sr$^{-1}$), divided by the peak value of $I_{\nu}$ at 7, 8.6, 11.3,
and 12.7 \micron\ IEFs are marked ({\it dotted vertical lines}).  The
difference between the rectified increasing and decreasing wavelength
scans of the source-sky spectra ({\it dotted line}) is plotted as an
estimate of the uncertainty.  \label{rplot} }
\end{figure}

We have divided the spectrophotometric images of vdB 17 and 59 into
radial bins, characterizing each bin by the incident UV starlight,
$G_0$.  We scale the values of $G_0$ in Table \ref{tab:gfac} by the
inverse square of the projected angular separation of each pixel from
the illuminating star.  Figure \ref{rplot} shows the average spectra
of different radial bins within vdB\,17 and vdB\,59.  The values of
$G_0$ in these bins are equally spaced in log($G_0$); they vary over
$G_0$ = 20 -- 4 $\times$ 10$^3$ for vdB\,17, and over $G_0$ = 200 -- 6
$\times$\ 10$^4$ for vdB\,59.

Near-infrared long-slit spectroscopy of vdB 17 and 59 (Martini,
Sellgren, \& DePoy 1999), as well as our unpublished ground-based
near-infrared imaging and the {\it ISO} images we analyze here,
suggest that vdB 17 has a much simpler geometry than vdB 59.  Thus our
use of the projected distance between a pixel and the central star to
estimate $G_0$ for each pixel is likely to be a better approximation
for vdB 17 than for vdB 59.

Figure \ref{rplot} shows interesting variations in the spectra of vdB
17 as a function of nebular position (or $G_0$).  At first glance, it
appears that the 8.6 $\mu$m IEF steadily becomes less prominent in the
spectra at low values of $G_0$.  A more careful examination, however,
shows that the {\it peak} intensity of the 8.6 $\mu$m IEF, relative to
the peak intensity of the 7.7 $\mu$m IEF, is relatively constant with
$G_0$.  Instead, it seems that the apparent weakening of the 8.6
$\mu$m IEF at low $G_0$ is rather due to a broadening of the
overlapping 7.7 $\mu$m IEF.

\begin{figure}
\figurenum{10}
\plotone{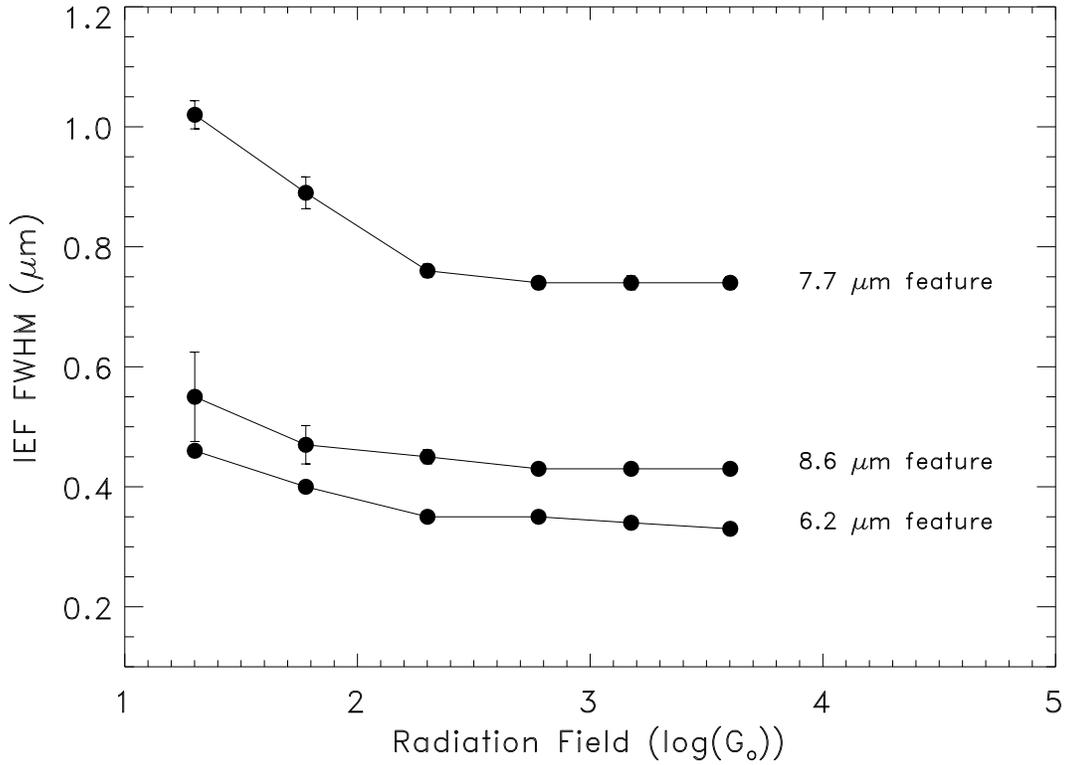}
\figcaption[]{Full-width at half-maximum (FWHM) of the 6.2 $\mu$m, 7.7
$\mu$m, and 8.6 $\mu$m IEFs vs.  incident UV intensity, log($G_0$),
for different nebular positions within vdB\,17.  FWHM were derived by
Lorentzian fits to the spectra shown in Fig.  \ref{rplot} (see text).
Error bars for each radial bin or annulus are derived from the
difference between spectra of the top half of the annulus and the
bottom half of the annulus.  \label{fwvsgo} }
\end{figure}

Figure \ref{fwvsgo} quantifies the change in the full-width at
half-maximum (FWHM) of the 7.7 $\mu$m IEF with $G_0$ in vdB 17.  We
adopted the method of Boulanger et al.  (1998a) to measure the FWHMs
of different IEFs, by fitting a linear baseline to continuum points
near 5 and 15 $\mu$m, and then fitting a blend of three Lorentzians to
the 6.2, 7.7, and 8.6 $\mu$m IEFs.  Figure \ref{fwvsgo} clearly shows
that the 7.7 $\mu$m IEF is systematically broader at low values of
$G_0$ in vdB 17.  Figure \ref{fwvsgo} also suggests that perhaps the
6.2 $\mu$m IEF may also broaden with lower $G_0$, but the evidence is
not as compelling as it is for the 7.7 $\mu$m IEF.  No change in the
FWHM of the 8.6 $\mu$m IEF is observed within the uncertainties.

The spectra of vdB 59, in contrast to those of vdB 17, do not show any
strong systematic dependence on $G_0$.  The width of the 7.7 $\mu$m
IEF and the contrast of the 8.6 $\mu$m IEF against the red wing of the
7.7 $\mu$m IEF show little change in Figure \ref{rplot}.  This may be
due to the more complex geometry of vdB 59, where emission from
regions close to and far from the star appear to contribute to the
same line of sight (Martini et al.  1999).  Alternatively, this may be
because $G_0$ in vdB 59 does not extend to the low values ($G_0$ = 20
-- 60) where the most marked spectral changes are observed in vdB 17.

The origin of the widths of the IEFs is quite controversial and this
complicates any explanation of why the 7.7 $\mu$m IEF width changes
spatially within vdB\,17.  An increase in PAH temperature should
broaden the IEFs, but that would also result in a shift in the IEF
central wavelengths (Joblin et al.  1995).  We, however, do not
observe any such shift in the 7.7 $\mu$m IEF central wavelength.  An
increase in particle size might broaden the IEFs (Boulanger et al.
1998).  A change in the mix of the IEF emitters, all of which have
slightly different wavelengths, could also lead to a change in the IEF
widths (Cook \& Saykally 1998; Le Coupanec et al.  1998; Papoular
1999).  Other broadening mechanisms are discussed by the above
authors; whichever mechanism accounts for the broadening of the 7.7
$\mu$m IEF that we observe in vdB\,17 needs to be consistent with the
much lower, or non-existent, broadening we observe in the 6.2 and 8.6
$\mu$m IEFs, and with the constancy in the central wavelengths of all
the observed IEFs.

The {\it peak} strengths of the 6.2 and 8.6 $\mu$m IEFs, relative to
the peak strength of the 7.7 $\mu$m IEF, show little variation among
spectra at different distances within vdB\,17 and vdB 59, over $G_0$ =
20 -- 6 $\times$ 10$^4$ (Fig.  \ref{rplot}).  The peak intensity of
the 8.6 $\mu$m IEF of vdB\,17, relative to the peak intensity of the
7.7 $\mu$m IEF, has values of 0.53 -- 0.56 in Figure \ref{rplot}.  A
similar constancy in peak intensity (relative to the peak 7.7 $\mu$m
IEF intensity) is observed for the 8.6 $\mu$m IEF within vdB 59 and
for the 6.2 $\mu$m IEF within both vdB 17 and vdB 59.

Figure \ref{rplot} also shows marginal evidence for a weak emission
feature near 6.9 $\mu$m appearing in the spectra of one or two of the
most distant radial bins within vdB 17.  A weak interstellar emission
feature at 6.9 $\mu$m was first discovered by Bregman et al.  (1983)
and further analyzed by Cohen et al.  (1986).  This emission feature
can be confused with [\ion{Ar}{2}] emission at 7.0 $\mu$m, and so can
only be reliably detected at low spectral resolution in sources free
of ionized hydrogen, such as reflection nebulae.  The 6.9 $\mu$m
feature is not detected in spectra of radial bins closer to the star
in vdB 17 (Fig.  \ref{rplot}).  The 6.9 $\mu$m feature is also not
detected in the spectra of any radial bin for vdB 59 (Fig.
\ref{rplot}), or in the spatially integrated {\it ISO} spectra of vdB
17 or vdB 59 (Fig.  \ref{iso2}), vdB 133 (Fig.  \ref{iso1}; Paper I),
vdB 139 (NGC 7023; D.  Cesarsky et al.  1996; Laureijs et al.  1996)
or vdB 106 ($\rho$ Oph; Boulanger et al.  1996).  The 0--0 S(5) H$_2$
line at 6.9 $\mu$m is unlikely to be a contributor, as other H$_2$
lines with similar predicted strengths, at 8.0, 9.7, and 12.3 $\mu$m,
are not detected, and because the 1-0 S(1) H$_2$ emission in our
unpublished narrowband images of vdB 17 is concentrated near the star
at higher G$_0$ values.  We are therefore cautious in claiming that we
have detected 6.9 $\mu$m emission in the outermost regions of vdB\,17.

\subsubsection{The 8.6 $\mu$m emission feature }

Laboratory data and theoretical calculations for PAHs show that the
8.6 to 11.3 $\mu$m IEF flux ratio, $I$(8.6 $\mu$m)/$I$(11.3 $\mu$m),
is much higher in ionized than neutral PAHs (De Frees et al.  1993;
Szczepanski \& Vala 1993; Hudgins et al.  1994; Pauzat et al.  1995,
1997; Langhoff 1996; Cook \& Saykally 1998; Allamandola et al.  1999).
The 8.6 $\mu$m and 11.3 $\mu$m bands in PAHs correspond to the
in-plane bend and out-of-plane bend, respectively, of the same
aromatic C--H bond.  Thus $I$(8.6 $\mu$m)/$I$(11.3 $\mu$m) should be a
useful measure of PAH ionization because it should be less sensitive
to PAH size or PAH hydrogenation than the ratio of an aromatic C--C
vibration to an aromatic C--H vibration.

The mean PAH ionization is predicted to increase with $G_0$ (Bakes \&
Tielens 1994, 1998; Salama et al.  1996; Dartois and d'Hendecourt
1997).  One test of the PAH hypothesis, therefore, is to measure
$I$(8.6 $\mu$m)/$I$(11.3 $\mu$m) as a function of $G_0$.  Joblin et
al.  (1996) have observed $I$(8.6 $\mu$m)/$I$(11.3 $\mu$m) to increase
closer to the star in the reflection nebula NGC 1333/SVS-3.  They
interpret this as due to PAHs becoming increasingly ionized near the
B-type illuminating star.

Figures \ref{lfit}, \ref{lrat}, and Table \ref{tab:linerat}, however,
demonstrate how strongly the measured value of $I$(8.6
$\mu$m)/$I$(11.3 $\mu$m) in our data depends on the method used to
separate the 8.6 $\mu$m IEF from other spectral components such as the
7.7 $\mu$m IEF and the 6 -- 9 $\mu$m broad emission bump.  A local
baseline (Method 2) to measure the strength of the 8.6 $\mu$m IEF
would lead us to conclude that the 8.6 $\mu$m IEF virtually disappears
at low values of $G_0$ in vdB 17, as can be seen qualitatively in
Figure \ref{rplot}.  The observed broadening of the 7.7 $\mu$m IEF at
low $G_0$ in vdB 17 (Figure \ref{fwvsgo}) will, however, cause an
increase in strength in the wing of the 7.7 $\mu$m IEF.  Because the
red wing of the 7.7 $\mu$m IEF defines the local continuum for the 8.6
$\mu$m IEF in Method 2, any broadening of the 7.7 $\mu$m IEF (at
constant 7.7 $\mu$m peak intensity) naturally results in an apparent
weakening in the 8.6 $\mu$m IEF strength measured by this method.
When we treat the 7.7 and 8.6 $\mu$m IEFs as a blend of two
symmetrical features lying above a much lower continuum level (Method
1), then we find that the 8.6 $\mu$m IEF shows little variation with
$G_0$ in vdB 17.

We conclude that $I$(8.6 $\mu$m)/$I$(11.3 $\mu$m) is extremely
dependent on how the 8.6 $\mu$m IEF is measured (Figs.  \ref{lfit} and
\ref{lrat}; Table \ref{tab:linerat}).  Joblin et al.  (1996) use a
local baseline (Method 2) in measuring the strength of the 8.6 $\mu$m
IEF, because their ground-based spectra do not extend to short enough
wavelengths to include the entire 7.7 $\mu$m IEF.  The increase in the
7.7 $\mu$m IEF width with decreasing $G_0$ we observe in vdB 17 (Figs.
\ref{rplot} and \ref{fwvsgo}) leads us to urge caution in using
$I$(8.6 $\mu$m)/$I$(11.3 $\mu$m) to quantify PAH ionization.

\subsubsection{The ratio of 
6 -- 10 $\mu$m emission to 10 -- 14 $\mu$m emission }

The difficulty of quantifying the observed 8.6 $\mu$m IEF strength
inspires us to search for a more robust way to test predictions of the
PAH ionization state as a function of $G_0$.  Laboratory data and
theoretical calculations for PAHs show that the sum of all the PAH
bands at 6 -- 10 $\mu$m, relative to the sum of all the PAH bands at
10 -- 14 $\mu$m, is at least an order of magnitude higher in ionized
PAHs than in neutral PAHs (De Frees et al.  1993; Szczepanski \& Vala
1993; Hudgins et al.  1994; Pauzat et al.  1995, 1997; Langhoff 1996;
Cook \& Saykally 1998; Allamandola et al.  1999).  We therefore have
measured $I$(5.50--9.75 $\mu$m)/$I$(10.25--14.0 $\mu$m), the ratio of
the integrated intensity at 5.50 -- 9.75 $\mu$m to the integrated
intensity at 10.25 -- 14.0 $\mu$m, in {\it ISO} spectra of reflection
nebulae, as an alternate way to search for the spectroscopic signature
of changing PAH ionization state.

\begin{figure}
\figurenum{11}
\plotone{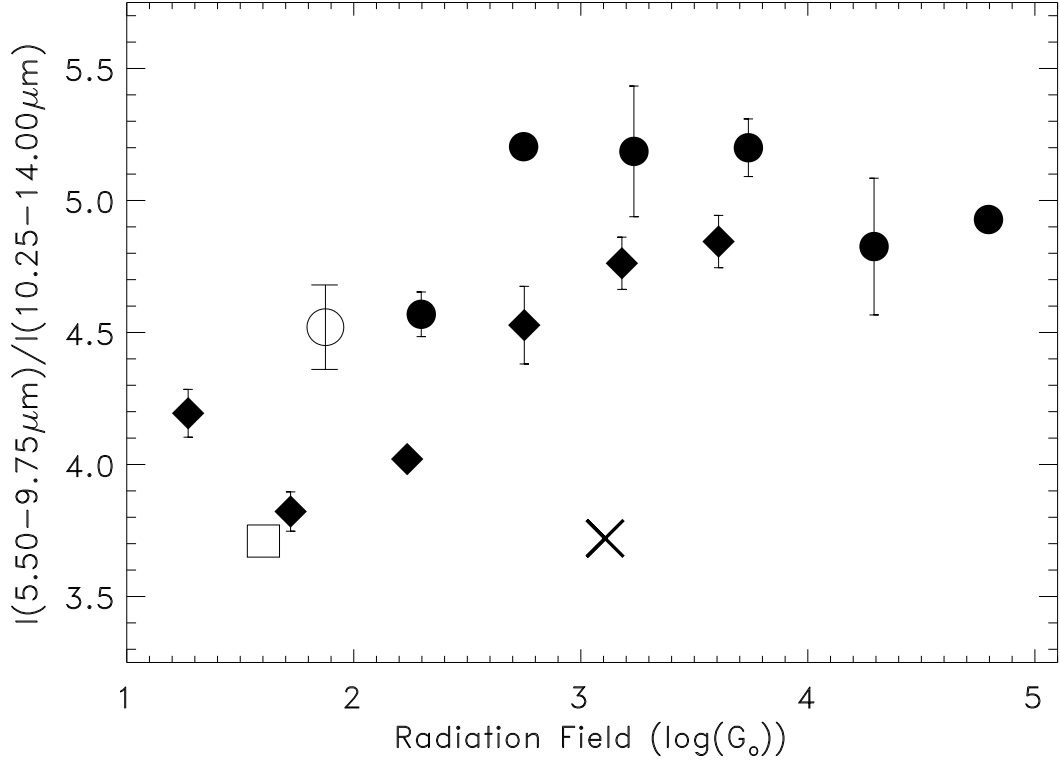}
\figcaption[]{Ratio of the integrated spectral intensity at 5.50 --
9.75 $\mu$m to the integrated spectral intensity at 10.25 -- 14.0
$\mu$m, $I$(5.50--9.75 $\mu$m)/$I$(10.25--14.0 $\mu$m), vs.  the
incident UV intensity, log($G_0$).  Ratios at different nebular
positions within vdB\,17 (NGC 1333; {\it filled diamonds}) and vdB\,59
(NGC 2068; {\it filled circles}) are derived from spectra in Fig.
\ref{rplot}.  Ratios for the spatially averaged spectra of vdB 133
({\it open circle}; Fig.  \ref{iso1}{\it b} and Paper I), vdB 106 =
$\rho$ Oph ({\it open square}; Boulanger et al.  1996), and vdB 139 =
NGC 7023 ({\it cross}; D.  Cesarsky et al.  1996) are also plotted.
Error bars for vdB 133 are derived from the difference between the
rectified increasing and decreasing wavelength scans of the source-sky
spectra.  Error bars for each radial bin or annulus in vdB 17 are
derived from the difference between spectra of the top half of the
annulus and the bottom half of the annulus.  Error bars for each
radial bin or annulus in vdB 59 are derived from the difference
between spectra of the right half of the annulus and the left half of
the annulus.  \label{splrat} }

\end{figure}

Figure \ref{splrat} presents our measurements of $I$(5.50--9.75
$\mu$m)/$I$(10.25--14.0 $\mu$m) as a function of $G_0$ in reflection
nebulae.  We subtracted a continuum from each spectrum before
measuring $I$(5.50--9.75 $\mu$m)/$I$(10.25--14.0 $\mu$m).  The 5.50 --
9.75 $\mu$m continuum was defined by a linear fit to data points near
5.3 and 10.0 $\mu$m.  The 10.25 -- 14.0 $\mu$m continuum was defined
by a linear fit to data points near 10.0 and 14.5 $\mu$m.  We plot a
single value of $I$(5.50--9.75 $\mu$m)/$I$(10.25--14.0 $\mu$m) for vdB
106 ($\rho$ Oph; Boulanger et al.  1996), vdB 133 (Fig.
\ref{iso1}{\it b} and Paper I), and vdB 139 (NGC 7023; D.  Cesarsky et
al.  1996).  For vdB 17 and vdB 59, we plot $I$(5.50--9.75
$\mu$m)/$I$(10.25--14.0 $\mu$m) at six values of $G_0$, derived from
the spectra of the radial bins shown in Figure \ref{rplot}.

The remarkable conclusion from Figure \ref{splrat} is that
$I$(5.50--9.75 $\mu$m)/$I$(10.25--14.0 $\mu$m) changes by no more than
40\% among sources with $G_0$ = 20 -- 6 $\times$ 10$^4$.  This
observational ratio should not depend at all on how the observed
interstellar spectra are divided into separate narrow IEFs and broader
emission components.  We emphasize that $I$(5.50--9.75
$\mu$m)/$I$(10.25--14.0 $\mu$m) should be a more robust and sensitive
measure of the PAH ionization state than $I$(8.6 $\mu$m)/$I$(11.3
$\mu$m).

Our observation that the mid-infrared spectra of vdB 17 and vdB 59 are
very similar for $G_0$ = 20 -- 6 $\times$ $10^4$ (Figs.  \ref{rplot}
and \ref{splrat}) are in strong agreement with other recent work.
Boulanger et al.  (1998b) concludes that the IEF flux ratios in {\it
ISO} spectra of the Chamaeleon cloud, the $\rho$ Oph cloud, NGC 7023,
and M17 show no systematic dependence on $G_0$, over $G_0$ = 1 --
10$^5$.  Onaka (1999) finds that the IEF flux ratios in {\it ISO}
spectra of the Carina nebula are independent of $G_0$, over $G_0$ =
100 -- 10$^5$.

\section{Conclusions}

We have obtained 5 -- 15 $\mu$m ISOCAM CVF spectroscopy of six
reflection nebulae, in order to test predictions of interstellar
models for the infrared emission features.  Our sample includes a
range of UV field hardness ($T_{\rm eff}$ = 3,600 K -- 19,000 K for
the illuminating stars; {$F_{(\lambda < 400 {\rm nm})}$/$F_{total}$}\
= 0.01 -- 0.87) and UV field intensity ($G_0$ = 20 -- 6 $\times$
10$^4$).

Our first goal was to search for spectroscopic differences as a
function of $T_{\rm eff}$ for the illuminating stars.  Detection of
such spectroscopic differences could potentially reconcile 12 $\mu$m
broad-band {\it IRAS} observations showing that $\Delta \nu I_\nu$(12
$\mu$m)/$I_{\rm bol}$(FIR) is independent of $T_{\rm eff}$, for
reflection nebulae illuminated by stars $T_{\rm eff}$ = 5,000 --
33,000 K (Sellgren et al.  1990), with model predictions that the {\it
IRAS} 12 $\mu$m emission is dominated by PAHs which contribute much
more to the total interstellar extinction curve at UV wavelengths than
at visible wavelengths (D\'esert et al.  1990; Joblin et al.  1992;
Siebenmorgen \& Kr\"ugel 1992; Dwek et al.  1997; Silva et al.  1998).

A second goal was to search for spectroscopic differences as a
function of $G_0$.  Interstellar models predict that more PAHs are
positively charged when $G_0$ increases (Bakes \& Tielens 1994, 1998;
Salama et al.  1996; Dartois and d'Hendecourt 1997).  Laboratory data
and theoretical calculations for PAHs show that $I$(8.6
$\mu$m)/$I$(11.3 $\mu$m) and $I$(5.50--9.75 $\mu$m)/$I$(10.25--14.0
$\mu$m) are both much higher in ionized than neutral PAHs (De Frees et
al.  1993; Szczepanski \& Vala 1993; Hudgins et al.  1994; Pauzat et
al.  1995, 1997; Langhoff 1996; Cook \& Saykally 1998; Allamandola et
al.  1999).

We detect {\it interstellar} IEFs at 6.2, 7.7, 8.6, 11.3, and 12.7
$\mu$m, broad emission features at 6 -- 9 $\mu$m and 11 -- 13 $\mu$m,
and continuum emission at 5 -- 15 $\mu$m in the reflection nebulae vdB
17, vdB 59, and vdB 133.  These nebulae were the brightest in our
sample and illuminated by hotter stars ({$F_{(\lambda < 400 {\rm
nm})}$/$F_{total}$}\ = 0.22 -- 0.87).  The normalized peak fluxes of
the different IEFs vary by less than a factor of two among these three
nebulae, with no systematic dependence on $T_{\rm eff}$.  No evidence
is seen among these spectra for disappearance of IEFs, appearance of
new spectral features, shifts in IEF wavelengths, or dramatic changes
in the strengths of IEFs relative to the broad emission bumps or the
continuum emission.

We do not detect any feature emission or continuum emission in vdB
101, vdB 111, and vdB 135.  These nebulae were the faintest in our
sample and were illuminated by cooler stars ({$F_{(\lambda < 400 {\rm
nm})}$/$F_{total}$}\ = 0.01 -- 0.20).  Their upper limits are
consistent with their low measured 12 $\mu$m {\it IRAS} surface
brightnesses.

We compare spatially averaged spectra of vdB 17, vdB 59, and vdB 133
to published spectra of vdB 106 ($\rho$ Oph; Boulanger et al.  1996)
and vdB 139 (NGC 7023; D.  Cesarsky et al.  1996).  Qualitative
comparison of spectra show all these spectra are very similar over a
range of UV hardness ($T_{\rm eff}$ = 6,800 -- 22,000; {$F_{(\lambda <
400 {\rm nm})}$/$F_{total}$}\ = 0.22 -- 0.92) and UV intensity ($G_0$
= 40 -- 1800).

We find that quantifying the IEF fluxes is quite sensitive to the
method by which the spectra are separated into IEFs, broad emission
bumps, and continuum.  For instance, $I$(8.6 $\mu$m)/$I$(11.3 $\mu$m)
shows no dependence on $T_{\rm eff}$ for any given method of defining
the IEF fluxes, but $I$(8.6 $\mu$m)/$I$(11.3 $\mu$m) shows strong
systematic differences between the values derived by different
methods.

We also quantify the contributions of IEFs, broad emission bumps, and
continuum to broad-band filters such as the {\it IRAS} 12 $\mu$m
filter and the ISOCAM LW\,2 (6.75 $\mu$m) filter.  Again, for any
given method of distinguishing between these separate spectral
components, the contributions of each component to the broad-band
filters are independent of $T_{\rm eff}$, but the absolute value of
the contribution depends strongly on the method for how the spectrum
is divided among IEFs, broad emission bumps, and continuum.

We divide the spectra of vdB 17 and vdB 59 into radial bins, with the
UV field of each bin characterized by $G_0$.  We find that the {\it
peak} intensities of the 6.2 and 8.6 $\mu$m IEFs, compared to the peak
intensity of the 7.7 $\mu$m IEF, show no variation over $G_0$ = 20 --
6 $\times$ 10$^4$ in vdB 17 and vdB 59.

In the most distant regions of vdB 17, we observe that the 8.6 $\mu$m
IEF appears weaker due to increased blending with the 7.7 $\mu$m IEF.
We quantify the FWHM of the 7.7 $\mu$m feature, and find that it grows
systematically broader at low $G_0$ in vdB 17.  We conclude that the
low contrast of the 8.6 $\mu$m IEF at low $G_0$ in vdB 17 is the
result of the increased width of the 7.7 $\mu$m IEF.  This observed
broadening of the 7.7 $\mu$m IEF, combined with the sensitivity of
$I$(8.6 $\mu$m)/$I$(11.3 $\mu$m) to baseline placement for the 8.6
$\mu$m IEF, suggests that caution should be used in inferring PAH
ionization from $I$(8.6 $\mu$m)/$I$(11.3 $\mu$m).

We suggest using $I$(5.50--9.75 $\mu$m)/$I$(10.25--14.0 $\mu$m) as an
alternate method of testing the predictions of PAH ionization as a
function of $G_0$.  We measure this in single nebular positions in vdB
106, vdB 133, and vdB 139 and at different nebular positions within
vdB 17 and vdB 59.  We observe that $I$(5.50--9.75
$\mu$m)/$I$(10.25--14.0 $\mu$m) decreases by no more than 40\% over
$G_0$ = 20 -- 6 $\times$ 10$^4$.

Our observational results are in strong agreement with Boulanger et
al.  (1998b), who concludes that the IEF flux ratios in {\it ISO}
spectra of the Chamaeleon cloud, the $\rho$ Oph cloud, NGC 7023, and
M17 show no systematic dependence on $G_0$ over $G_0$ = 1 -- 10$^5$.
Our results also strongly support Onaka (1999), who finds that the IEF
flux ratios in {\it ISO} spectra of the Carina nebula are independent
of $G_0$ over $G_0$ = 100 -- 10$^5$.

Much is yet to be gleaned from this {\it ISO} survey of reflection
nebulae and especially the detection of IEFs toward vdB\,133.  In a
future paper we will present a quantitative analysis of the IEF energy
budget using laboratory absorption profiles of various carrier
candidates (neutral PAHs, ionized PAHs, amorphous hydrocarbons, and
coal tars).  We will also present follow-up 7.75 \micron\ (LW\,6) {\it
ISO} observations which cover an 8$'$ by 8$'$ mosaic around vdB\,133,
detailing the spatial distribution of the surrounding IEF emission.
The relation of the IEF carriers to the gas and other components of
dust surrounding vdB\,133 will be explored by comparing the {\it ISO}
images to ground-based infrared and molecular line ($^{12}$CO J=2--1)
data that have also been obtained.

\acknowledgments

We would like to thank A.  L\'eger, Jean-Loup Puget and C.  Moutou for
hosting us at the Institut d'Astrophysique Spatiale.  We are grateful
to W.  Reach, D.  Cesarsky and F.  Boulanger for providing their
expertise and invaluable help in the reduction of our ISOCAM data.  We
appreciate the generosity of D.  Cesarsky and colleagues for providing
us with their ISOCAM spectrum of NGC 7023, and F.  Boulanger and
colleagues for providing us with their ISOCAM spectrum of $\rho$ Oph.

We acknowledge NASA support of the {\it ISO} data analysis through NAG
5--3366 and JPL contract 961562, and also support from NATO
Collaborative Research Grant 951347.  This work was carried out in
part at the Jet Propulsion Laboratory, California Institute of
Technology, under contract with the National Aeronautics and Space
Administration.

% ZZ-end-of-text-zz

%-----------------------TABLE 1 -----------------------
 \makeatletter
 \def\jnl@aj{AJ}
 \ifx\revtex@jnl\jnl@aj\let\tablebreak=\nl\fi
 \makeatother
 \begin{deluxetable}{llllllll}
 \scriptsize
 \tablewidth{0pc}
 \tablecaption{Stellar Data \label{tab:stelpar}}
 \tablehead{
     \colhead{Nebula$^a$} 
   & \colhead{Star\phantom{xxxxxxxxxx}} 
   & \colhead{Spectral Type\phantom{XXX}} 
   & \colhead{$T_{\rm eff}$\phantom{xxxxxxxx}}  
   & \colhead{${F{(\lambda < 400 {\rm nm})}} \over {F{total}}$}
   & \colhead{${F{(\lambda < 240 {\rm nm})}} \over {F{total}}$}
   & \colhead{${\Delta \nu I_\nu (12 \mu {\rm m})^b} \over {I_{\rm bol}(FIR)}$}
   & \colhead{Refs$^g$} \nl
 }

 \startdata
 \multicolumn{3}{l}{Observed Sources$^c$} & & & &  \nl
 \cline{1-2}  

 vdB\,17 & BD\,+30$\;$549  & 
 B8$\;$V    & 11,000 & 0.55 & 0.33 & 0.09 & 1 \nl

 vdB\,59     & HD\,38563B  & 
 B2$\;$II--III  & 19,000 & 0.87  & 0.69 & 0.14 & 2  \nl

 vdB\,101 & HD\,146834  & 
 G5$\;$III   & 5,000  & 0.07 & 0.003 & 0.29 & 1 \nl

 vdB\,111 & HD\,156697  & 
 F0$\;$n     & 7,300  & 0.15 -- 0.20$^d$ & 0.02 -- 0.03 & 0.21 & 1 \nl

 vdB\,133 & HD\,195593(A $+$ B)  & 
               F5$\;$Iab $+$ B7$\;$II 
             & 6,800 $+$ 12,000 & 0.22$^e$ & 0.09$^e$ & 0.19 & 1,3 \nl

 vdB\,135 & BD\,+31$\;$4152  & 
 M1$\;$IIIe  & 3,600  & 0.01 & 0.00005 & 0.11 & 1  \nl

 \multicolumn{3}{l}{Comparison Sources} & & & & \nl
 \cline{1-2}

 vdB\,139 & HD\,200775  & 
 B3$\;$IVe    & 17,000 & 0.84 & 0.65 & 0.09 & 1,4 \nl

 vdB\,106 & HD\,147933 $+$ HD\,147934 & 
      B2$\;$IV-V $+$ B2$\;$IV-V & 22,000 $+$ 22,000 & 0.92$^e$ 
      & 0.77$^f$ & $\cdots$ &  5  \nl
 \enddata

\tablenotetext{a}{vdB\,17 = NGC\,1333,  vdB\,59 = NGC\,2068, 
                  vdB\,106 =$\rho$\,Oph, and vdB\,139 = NGC\,7023.}
\tablenotetext{b}{Based on data from Tables 2 and 3 in 
Sellgren et al. 1990.  The adopted width of the 12 $\mu$m IRAS 
                  bandpass is 1.75 $\times$ 10$^{13}$ Hz.}
\tablenotetext{c}{Sources observed in this study}
\tablenotetext{d}{The range in 
{$F_{(\lambda < 400 {\rm nm})}$/$F_{total}$}\ 
and {$F_{(\lambda < 240 {\rm nm})}$/$F_{total}$}\  
stems from the uncertainty 
                  in the luminosity class of vdB\,111.}
\tablenotetext{e}{
{$F_{(\lambda < 400 {\rm nm})}$/$F_{total}$}\ 
for vdB\,133 is equal to the sum of {$F_{(\lambda < 400 {\rm nm})}$}\  
for stars A and B divided by the sum of $F_{total}$\  for stars A and B.
{$F_{(\lambda < 240 {\rm nm})}$/$F_{total}$}\ 
for vdB\,133 is equal to the sum of {$F_{(\lambda < 240 {\rm nm})}$}\  
for stars A and B divided by the sum of $F_{total}$\  for stars A and B.}
\tablenotetext{f}{
{$F_{(\lambda < 400 {\rm nm})}$/$F_{total}$}\ 
for $\rho$ Oph is equal to the sum of {$F_{(\lambda < 400 {\rm nm})}$}\ 
for stars HD\,147933 and HD\,147934 divided 
by the sum of $F_{total}$\  for stars HD\,147933 and HD\,147934.
$F_{(\lambda < 240 {\rm nm})}$/$F_{total}$\ 
for $\rho$ Oph is equal to the sum of $F_{(\lambda < 240 {\rm nm})}$\  
for stars HD\,147933 and HD\,147934 divided 
by the sum of $F_{total}$\  for stars HD\,147933 and HD\,147934.}
\tablenotetext{g}{Spectral classification references: 
                  (1) Racine 1968;
                  (2) Strom et al. 1975;
                  (3) Uchida, Sellgren, \& Werner 1998 (Paper I); \\
 	          (4) Witt \& Cottrell 1980;                   
                  (5) de Geus, de Zeeuw, \& Lub 1989.
                 }

 \end{deluxetable}
 \normalsize
%-----------------------TABLE 2-----------------------
\makeatletter
\def\jnl@aj{AJ}
\ifx\revtex@jnl\jnl@aj\let\tablebreak=\nl\fi
\makeatother
\begin{deluxetable}{lrr}
\tablewidth{155mm}
\tablecaption{Observation Sequences \label{tab:obswave}}
\tablehead{\colhead{Sequence} & Wavelengths ($\micron$)$^a$  
& Observed Sources (vdB) \nl}
\startdata

1a & \phn5.8, \phn6.2, \phn7.0, \phn9.5\phantom{$^b$} 
   & \phn101, \phn133, \phn135 \nl
1b & \phn4.5$^b$, \phn6.2, \phn7.0, \phn9.5\phantom{$^b$} 
   & \phn111         \nl
2  &          \phn7.0, \phn7.6, \phn8.4\phantom{$^b$} 
   & \phn101, \phn111, \phn133, \phn135 \nl 
3  &          \phn8.4, \phn8.62, \phn9.5\phantom{$^b$} 
   & \phn101, \phn111, \phn133	\nl
4  &    10.5,    11.22,    12.0,    15.0$^c$	      
   & \phn101, \phn111, \phn133, \phn135 \nl

\enddata
\tablenotetext{a}{Unless otherwise noted all wavelengths refer to 
circular variable filter settings, which have a spectral resolution 
$\lambda / \Delta \lambda$ = 40.}
\tablenotetext{b}{Narrowband filter LW1 ($\Delta \lambda$ = 1 $\mu$m). }
\tablenotetext{c}{Narrowband filter LW9 ($\Delta \lambda$ = 2 $\mu$m). }

\end{deluxetable}

%-----------------------TABLE 3----------------------- 
 \makeatletter
 \def\jnl@aj{AJ}
 \ifx\revtex@jnl\jnl@aj\let\tablebreak=\nl\fi
 \makeatother 
 \begin{deluxetable}{lllllll}
 \scriptsize
 \tablewidth{35pc}
 \tablecaption{Field Positions \label{tab:obspos}}
 \tablehead{
     \multicolumn{1}{l}{Nebula}
   & \multicolumn{2}{l}{Stellar Position}
   & \multicolumn{2}{l}{Source offset from star}
   & \multicolumn{2}{l}{Sky offset from star} \nl
     \colhead{}
   & \colhead{ RA$\,$(2000)} 
   & \colhead{Dec$\,$(2000)}
   & \colhead{$\Delta$RA$\,(\,''\,)$} 
   & \colhead{$\Delta$Dec$\,(\,''\,)$}
   & \colhead{$\Delta$RA$\,(\,''\,)$} 
   & \colhead{$\Delta$Dec$\,(\,''\,)$} \nl
 }

 \startdata

 \multicolumn{3}{l}{Position Switched Observations} &  & \nl
 \cline{1-3}  

 vdB\phn101 & 16$\;$19$\;$07.5 & $-$20$\;$13$\;$03.9 & 0 & 0 
 & $+$562 & $-$112 \nl

 vdB\phn111 & 17$\;$18$\;$52.7 & \phs06$\;$05$\;$07.1 & 0 & 0 
 & $+$498 & $+$160 \nl

 vdB\phn133 & 20$\;$30$\;$59.1 & \phs36$\;$56$\;$09.2 & 0 & 0 
 & $-$1187 & $+$266 \nl

 vdB\phn135 & 20$\;$36$\;$45.5 & \phs32$\;$27$\;$17.3 & 0 & 0 
 & $-$1218 & $-$97 \nl

 &  &  &  & \nl

 \multicolumn{3}{l}{Full CVF Scan Observations}  &  &  \nl
 \cline{1-3}  

 vdB\phn17 (NGC\,1333)   & 03$\;$29$\;$19.8 
                         & \phs31$\;$24$\;$56.0& $+$12 & $-$26 
                         & $+$493  & $+$12  \nl

 vdB\phn59 (NGC\,2068)  & 05$\;$46$\;$44.6 
                        & \phs00$\;$05$\;$22.6 & $+$92 & $+$12  
                        & $+$1026 & $+$13  \nl

 vdB\phn101 & 16$\;$19$\;$07.5 & $-$20$\;$13$\;$03.9 & $+$39 & $+$20 
 & $+$470   & $-$24  \nl

 vdB\phn133 & 20$\;$30$\;$59.1 & \phs36$\;$56$\;$09.2 & $+$134 & $+$67 
 & $+$704   & $+$102 \nl

 vdB\phn135 & 20$\;$36$\;$45.5 & \phs32$\;$27$\;$17.3 & $-$96 & $+$201 
 & $-$771   & $-$86 \nl
 
\enddata

\end{deluxetable}
\normalsize

%-----------------------TABLE 4 -----------------------
 \makeatletter
 \def\jnl@aj{AJ}
 \ifx\revtex@jnl\jnl@aj\let\tablebreak=\nl\fi
 \makeatother
 \begin{deluxetable}{lrrrrrr}
 \scriptsize
 \tablewidth{0pc}
 \tablecaption{The Integration Regions \label{tab:intreg}}
 \tablehead{
     \colhead{Nebula\phantom{xxxxxxxxx}} 
   & \multicolumn{2}{l}{Region Center$^a$} 
   & \multicolumn{2}{l}{Region Size} 
   & \colhead{PA$^b$} 
   & \colhead{Comments$^c$} \nl     
     \colhead{ }
   & \colhead{$\Delta$RA($''$)}
   & \colhead{$\Delta$Dec($''$)}   
   & \colhead{$\Delta$X($''$)}   
   & \colhead{$\Delta$Y($''$)}   
   & \colhead{($\arcdeg$)}
   & \colhead{ } \nl
 }

\startdata

\multicolumn{3}{l}{Position Switched Observations}  &  &  &  &  \nl
\cline{1-3}
vdB\,101  & 0 & 0 
          & 150  & 150  &  9.4   & ER, XS        \nl
vdB\,111  & 0 & 0 
          & 150  & 150  & 85.9 & ER, XS        \nl
vdB\,133  & $+$72 & $+$46 
          & 66   & 30   & 20.7 & LE, XS        \nl
vdB\,135  & 0 & 0  
          & 150  & 150  & 5.6 & ER, XS        \nl

 & & & & & & \nl

\multicolumn{3}{l}{Full CVF Scan Observations}  &  &  &  &  \nl
\cline{1-3}
vdB\,17 (NGC\,1333)  & $+$14  & $-$22
          & 60  & 60  & 77.7   & LE            \nl
vdB\,59 (NGC\,2068)  & $+$90 & $+$28
          & 72  & 78  &  5.8   & LE            \nl
vdB\,101  & $+$34  & $+$227  
          & 192 & 192 & 10.1   & ER            \nl
vdB\,133  & $+$85 & $+$43
          & 60  & 48  & 83.8  & LE            \nl
vdB\,135  & $-$104  & $+$204
          & 192 & 192 & 65.5   & ER            \nl
\enddata

\tablenotetext{a}{ offset of the integration region center from the 
                   central star.}
\tablenotetext{b}{ The rotation of the Y-axis east of celestial north. }
\tablenotetext{c}{ Describes the subregion used to produce the 
                   spectral averages: LE --- containing localized IEF 
                   emission, ER --- the entire usable portion of the 
                   image, XS --- excluding the emission from the central 
                   star. }
\tablecomments{Columns 2 and 3 give the location of
the subregion center in $\Delta$RA, $\Delta$Dec
offsets from the illuminating star(s), columns 4 and 5 its $\Delta$X and
$\Delta$Y dimensions, respectively, and column 6 the position angle of
its Y-axis with respect to celestial north.  Column 7 provides the brief
description of the region, whether it encompasses the entire usable area
of the array (ER --- entire region), or a localized patch of IEF emission
(LE --- localized emission). In those cases where the image contained the
illuminating star (the position switched data), the star and the immediate
region containing a spurious arc-like reflection feature from the star
(typically pixels X,Y = [6--20, 11--24] of the array) were excluded (XS;
excluded star) from the spatial integration.}

\end{deluxetable}

%-----------------------TABLE-5----------------------
 \makeatletter
 \def\jnl@aj{AJ}
 \ifx\revtex@jnl\jnl@aj\let\tablebreak=\nl\fi
 \makeatother
 \begin{deluxetable}{lllll}
 \scriptsize
 \tablewidth{145mm}
 \tablecaption{Nebular Illumination \label{tab:gfac}}
 \tablehead{
     \colhead{Nebula\phantom{xxxxxxxxxxxxx}} 
   & \colhead{Nebula distance} 
   & \colhead{$\theta_{Neb}^a$\phantom{xxx} } 
   & \colhead{$G_0$ (91 nm $< \lambda < 240$ nm)$^b$ } 
   & \colhead{Refs$^c$}   \nl
     \colhead{ } 
   & \colhead{(pc) \phantom{xxxxxxxxx}} 
   & \colhead{(arcsec)} 
   & \colhead{ }  
   & \colhead{ }      \nl
   }
 \startdata

 \multicolumn{2}{l}{Observed Sources} \nl
 \cline{1-2}  

 vdB\,17 (NGC\,1333)        & 500   & 29    & 175   & 1       \nl

 vdB\,59 (NGC\,2068)        & 550   & 93    & 520   & 2       \nl

 vdB\,133                   & 1500  & 100   &  75   & 3, 4    \nl

 \multicolumn{2}{l}{Observed Sources} \nl
 \cline{1-2}  

 vdB\,139 (NGC\,7023)       & 440   & 43    & 1820   & 5       \nl

 vdB\,106 ($\rho$ Oph)      & 125   & 2400  & 40     & 6       \nl

 \enddata
 \tablenotetext{a}{The angular distance between illuminating star and 
                   the observed nebular region.}
 \tablenotetext{b}{$G_0$, the UV (91\,nm $< \lambda\ <$ 240 nm) 
                   radiation field at the nebula, due to the 
                   illuminating stars, in units of
                   1.6\,$\times$\,10$^{-3}$ ergs s$^{-1}$ cm$^{-2}$, 
                   the local interstellar radiation field as 
                   determined by Habing 1968}
 \tablenotetext{c}{ 
                   Distance references: 
                   (1) Strom, Grasdalen, \& Strom 1974;
                   (2) Racine 1968;                    
                   (3) Humphreys 1978; 
                   (4) Garmany \& Stencel 1992;  
                   (5) Whitcomb et al. 1981;
                   (6) de Geus, de Zeeuw, \& Lub 1989. 
                  }

 \end{deluxetable}
 \normalsize

%-----------------------TABLE-6----------------------
\makeatletter
\def\jnl@aj{AJ}
\ifx\revtex@jnl\jnl@aj\let\tablebreak=\nl\fi
\makeatother
\begin{deluxetable}{lllllll}
\scriptsize
\tablewidth{155mm}
\tablecaption{Normalized Integrated Line Fluxes \label{tab:linerat}}
\tablehead{ 
           \colhead{ } 
         & \colhead{ } 
         & \multicolumn{3}{c}{Observed Sources} 
         & \multicolumn{2}{c}{Comparison Sources} \nl
           \cline{3-5} 
           \cline{6-7}                 \nl
           \colhead{ }
         & \colhead{ }
         & \colhead{vdB\,17$^a$}   
         & \colhead{vdB\,59} 		
         & \colhead{vdB\,133}   
         & \colhead{vdB\,139}
         & \colhead{vdB\,106}          \nl	
          }

\startdata

\multicolumn{2}{c}{Method 1 Fluxes$^b$}  \nl
\cline{1-2} \nl
IEFs & (6.2 $\micron$)/(11.3 $\micron$)
     & 1.4  & 1.7  & 1.0  & 1.1  & 1.2          \nl
     & (7.7 $\micron$)/(11.3 $\micron$)    
     & 3.8  & 4.4  & 4.8  & 3.4  & 4.1          \nl
     & (8.6 $\micron$)/(11.3 $\micron$)    
     & 2.0  & 2.3  & 2.9  & 1.8  & 2.0          \nl
     & (12.7 $\micron$)/(11.3 $\micron$)   
     & 0.18 & 0.30 & 0.29 & 0.21 & 0.21         \nl 

Broad Bumps & (6 -- 9 $\micron$)/(11.3 $\micron$)   
            & 3.0  & 2.7  & 2.6  & 2.5  & 2.7   \nl
            & (11 -- 13 $\micron$)/(11.3 $\micron$) 
            & 1.6  & 1.2  & 1.1  & 1.4  & 1.8   \nl

Continuum   & (5.2 -- 15.1 $\micron$)/(11.3 $\micron$) 
            & 6.1  & 4.6  & 7.1  & 5.0  & 6.5   \nl

 & & & & & \nl

\multicolumn{2}{c}{Method 2 Fluxes} \nl
\cline{1-2} \nl
IEFs       & (6.2 $\micron$)/(11.3 $\micron$)   
           & 1.2  & 1.7  & 1.3   & 1.2   & 1.4       \nl
           & (7.7 $\micron$)/(11.3 $\micron$)
           & 1.7   & 2.0   & 2.3   & 2.0   & 2.0     \nl
           & (8.6 $\micron$)/(11.3 $\micron$)    
           & 0.28 & 0.44 & 0.29 & 0.24 & 0.36        \nl
           & (12.7 $\micron$)/(11.3 $\micron$)   
           & 0.18 & 0.18 & 0.24 & 0.30 & 0.23        \nl
Continuum  & (5.2 -- 15.1 $\micron$)/(11.3 $\micron$) 
           & 14.0 & 11.0 & 15.0 & 13.0 & 15.0   \nl
           & & & & & \nl

\multicolumn{2}{c}{Method 3 Fluxes}  \nl
\cline{1-2} \nl
IEFs        & (6.2 $\micron$)/(11.3 $\micron$)    
            & 1.4  & 1.8  & 1.0  & 1.2  & 1.0     \nl
            & (7.7 $\micron$)/(11.3 $\micron$)    
            & 2.3  & 3.0  & 2.4  & 2.5  & 2.1     \nl
            & (8.6 $\micron$)/(11.3 $\micron$)    
            & 1.4  & 1.7  & 1.7  & 1.3  & 1.3     \nl
            & (12.7 $\micron$)/(11.3 $\micron$)   
            & 0.87 & 0.50 & 0.50 & 0.63 & 0.60    \nl
Continuum   & (5 -- 15 $\micron$)/(11.3 $\micron$)  
            & 3.6  & 3.7  & 2.3  & 3.1  & 2.6     \nl
 & & & & & \nl

\multicolumn{2}{c}{11.3 $\micron$ Line Fluxes$^c$}            \nl
\cline{1-2} \nl
Method 1 (11.3 $\micron$)  &  
            & 1.3e$-3$ & 2.8e$-3$ & 1.5e$-4$ & $\cdots^d$  
            & 3.3e$-4$  \nl
Method 2 (11.3 $\micron$)  &  
            & 1.3e$-3$ & 3.2e$-3$ & 1.5e$-4$ & $\cdots^d$  
            & 3.3e$-4$  \nl
Method 3 (11.3 $\micron$)  &  
            & 2.2e$-3$ & 4.0e$-3$ & 3.2e$-4$ & $\cdots^d$  
            & 6.9e$-4$  \nl

\nl
\nl
\nl

\multicolumn{2}{c}{Total Band Fluxes}            \nl 
\cline{1-2} \nl
IRAS 12 $\micron^e$  &  
            & 1.1e$-2$ & 2.0e$-2$ & 1.4e$-3$ & $\cdots^g$  
            & 3.0e$-3$   \nl
ISO LW\,2$^f$        &  
            & 7.6e$-4$ & 1.7e$-3$ & 9.6e$-5$ & $\cdots^g$  
            & 1.9e$-4$    \nl

\tablenotetext{a}{vdB\,17 = NGC\,1333, vdB\, 59 = NGC\,2068, 
                  vdB\,139 = NGC\,7023, and vdB\, 106 = $\rho$\ Oph.}
\tablenotetext{b}{Integrated line flux from listed spectral feature 
                  (ergs s$^{-1}$ cm$^{-2}$ sr$^{-1}$) divided by the 
                  integrated flux within the 11.3 $\mu$m feature.} 
\tablenotetext{c}{Total integrated 11.3 $\mu$m line flux in units of 
                  ergs s$^{-1}$ cm$^{-2}$ sr$^{-1}$.}
\tablenotetext{d}{No integrated 11.3 $\mu$m line flux of vdB\,139 is 
                  presented since only a scaled version of the spectrum 
                  was available to us.}
\tablenotetext{e}{Total integrated source flux (ergs s$^{-1}$ cm$^{-2}$ 
                  sr$^{-1}$) expected in 12 $\micron$ IRAS band, 
                  found by the convolution of the source CVF spectrum 
                  with the 12 $\micron$ IRAS bandpass.}
\tablenotetext{f}{Total integrated source flux (ergs s$^{-1}$ cm$^{-2}$ 
                  sr$^{-1}$) expected in ISO LW\,2 band, found by the 
                  convolution of the source CVF spectrum with the ISO 
                  LW\,2 bandpass.}
\tablenotetext{g}{No integrated 12 $\mu$m line flux of vdB\,139 is presented 
                  since only a scaled version of the spectrum was 
                  available to us.}

\enddata
\end{deluxetable}
\normalsize

%-----------------------TABLE 7----------------------
\makeatletter
\def\jnl@aj{AJ}
\ifx\revtex@jnl\jnl@aj\let\tablebreak=\nl\fi
\makeatother
\begin{deluxetable}{lll}
\scriptsize
\tablewidth{145mm}
\tablecaption{Normalized Bandpass Fluxes \label{tab:lineband}}
\tablehead{\colhead{ }  
         & \multicolumn{2}{c}{Filter}   \nl
           \cline{2-3} 			\nl 
           \colhead{ }
         & \colhead{IRAS 12 $\micron$}   
         & \colhead{ISO LW 2 }          \nl	
          }

\startdata

Method 1 Fluxes    &                &             \nl
\cline{1-1}                                       \nl
IEFs/(Total$^a$)   & 0.35 -- 0.51$^b$             & 0.52 -- 0.59  
\phantom{$^b$}  \nl
Broad Bumps/Total  & 0.10 -- 0.18\phantom{$^b$}   & 0.22 -- 0.26  
\phantom{$^b$}  \nl
Continuum/Total    & 0.33 -- 0.45\phantom{$^b$}   & 0.18 -- 0.22  
\phantom{$^b$}  \nl

& & \nl

Method 2 Fluxes    &                &              \nl
\cline{1-1}                                        \nl
IEFs/Total         & 0.18 -- 0.26\phantom{$^b$}   & 0.26 -- 0.35 
\phantom{$^b$}  \nl
Continuum/Total    & 0.74 -- 0.82\phantom{$^b$}   & 0.65 -- 0.73 
\phantom{$^b$}  \nl
 
& & \nl

Method 3 Fluxes    &                &              \nl
\cline{1-1} \nl
IEFs/Total         & 0.58 -- 0.67\phantom{$^b$}   & 0.64 -- 0.71 
\phantom{$^b$}  \nl
Continuum/Total    & 0.29 -- 0.47\phantom{$^b$}   & 0.15 -- 0.22 
\phantom{$^b$}  \nl

\enddata
\tablenotetext{a}{Integrated flux of the spectral feature (erg cm$^{-2}$ 
s$^{-1}$ sr$^{-1}$) divided by the integrated flux within the broadband 
filter (IRAS 12 $\mu$m or ISO LW\,2) found by the convolution of the 
source CVF spectrum with the bandpass function.}
\tablenotetext{b}{The range of values presented represent the 
minimum and maximum from among the five sources, vdB\,17 (NGC\,1333), 
vdB\,59 (NGC\,2068), vdB\,133, vdB\,139 (NGC\,7023), vdB\,106 ($\rho$ Oph).}
  
\normalsize
\end{deluxetable}

%-----------------------------------------------------

\newpage

\end{document}